\documentclass[twocolumn,a4paper]{article} 
\pdfoutput=1

\usepackage[margin=15mm]{geometry}
\usepackage[hyphens]{url}
\usepackage{lmodern,microtype}
\usepackage{graphicx}
\usepackage{amsmath}
\usepackage{cite}
\usepackage{hyperref}
\hypersetup{colorlinks=true, linkcolor=black, citecolor=black,
  urlcolor=black}
\newcommand{\litref}[1]{Ref.~\cite{#1}}
\newcommand{\figref}[1]{Fig.~\ref{#1}}
\newcommand{\secref}[1]{Sec.~\ref{#1}}
\newlength{\breite}
\title{The Turbulent Nature of the Atmospheric Boundary Layer and its
  Impact on the Wind Energy Conversion Process}

\author{%
  Matthias W{\"a}chter, \and
  Hendrik Hei{\ss}elmann, \and
  Michael H{\"o}lling, \and
  Allan Morales, \and
  Patrick Milan, \and
  Tanja M{\"u}cke, \and
  Joachim Peinke, \and
  Nico Reinke, \and
  Philip Rinn%
  \thanks{%
    ForWind Center for Wind Energy Research,
    Institute of Physics,
    Carl von Ossietzky University,
    26111 Oldenburg, Germany, email: matthias.waechter@uni-oldenburg.de}%
}%

\date{\small Accepted for publication in the Journal of Turbulence on
  May 17, 2012.}

\begin{document}
\maketitle

\begin{abstract}
  Wind turbines operate in the atmospheric boundary layer, where they
  are exposed to the turbulent atmospheric flows. 
  As the response time of wind turbine is typically in the range of
  seconds, they are affected by the small scale intermittent
  properties of the turbulent wind.
  Consequently, basic features which are known for small-scale
  homogeneous isotropic turbulence, and in particular the well-known
  intermittency problem, have an important impact on the wind
  energy conversion process.
  We report on basic research results concerning the small-scale
  intermittent properties of atmospheric flows and their impact on
  the wind energy conversion process.
  The analysis of wind data shows strongly intermittent statistics
  of wind fluctuations. To achieve numerical modeling a
  data-driven superposition model is proposed.
  For the experimental reproduction and adjustment of intermittent
  flows a so-called active grid setup is presented. Its ability is shown to
  generate reproducible properties of atmospheric flows on the smaller
  scales of the laboratory conditions of a wind tunnel. 
  As an application example the response dynamics of different
  anemometer types are tested.
  To achieve a proper understanding of the impact of intermittent
  turbulent inflow properties on wind turbines we present methods of
  numerical and stochastic modeling, and compare the results to
  measurement data.
  As a summarizing result we find that atmospheric turbulence imposes
  its intermittent features on the complete wind energy conversion
  process. Intermittent turbulence features are not only present in
  atmospheric wind, but are also dominant in the loads on the turbine,
  i.e.{} rotor torque and thrust, and in the electrical power output
  signal. We conclude that profound knowledge of turbulent statistics
  and the application of suitable numerical as well as experimental
  methods are necessary to grasp these unique features and quantify
  their effects on all stages of wind energy conversion.
\end{abstract}


\section{Introduction}
\label{sec:intro}

Wind Energy is one of the promising renewable energies of the
presence, see e.g.~\litref{Lu2009a}. It is therefore of primary
importance for future sustainable energy supply systems.
During the past decades, considerable advances in science and engineering
have lead to wind energy converters (WECs) reaching efficiencies and
sizes which appeared impossible only a few years ago.

\begin{figure}
  \centering
  \includegraphics[height=50mm]{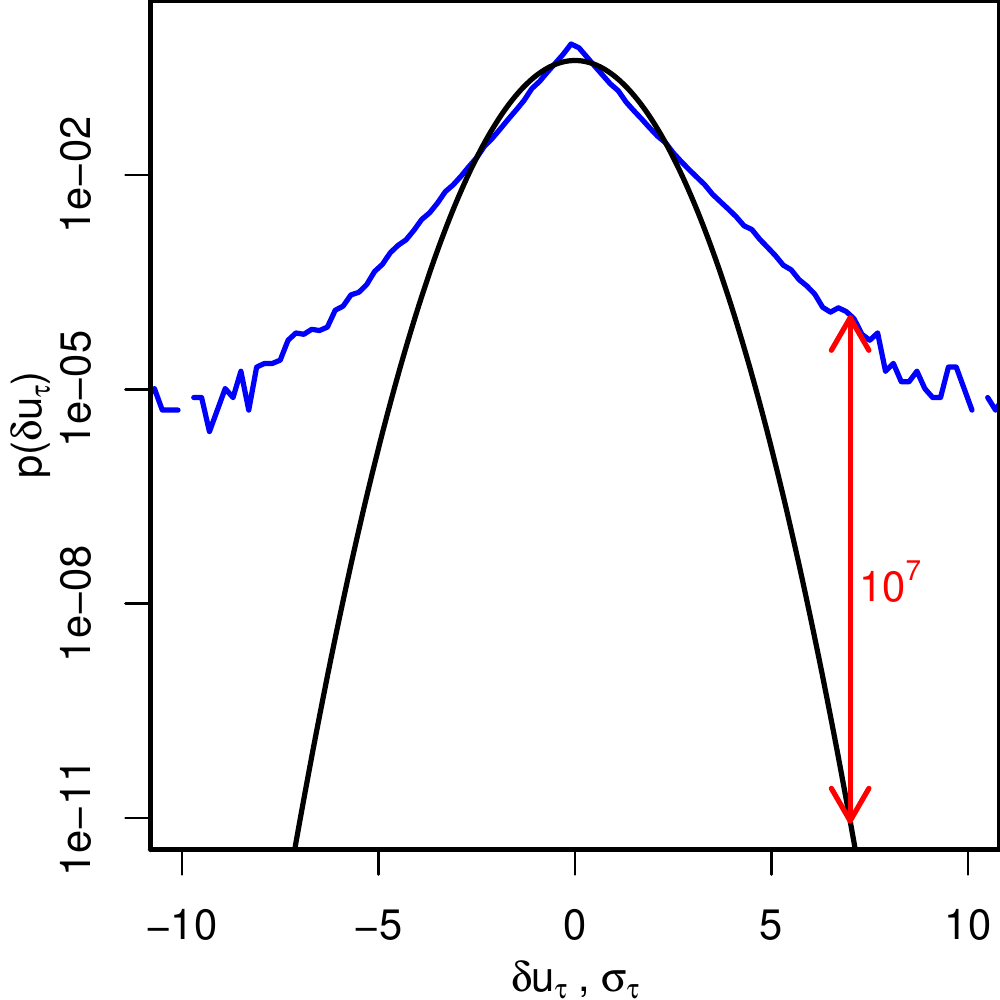}
  \caption{PDF of measured wind speed increments $\delta
    u_\tau(t)=u(t+\tau)-u(t)$ for time lag $\tau=3s$ (blue line)
    compared to a Gaussian PDF with same standard deviation (black
    line). At $7\sigma$ both differ by a factor of $10^7$. For the
    mean wind speed of 6.6\,m/s, $\tau=3s$ corresponds to a spatial
    separation of about 20\,m, which is a relevant scale for, e.g., rotor
    blades of current wind turbines.
    Note the logarithmic scale of the vertical axis. For details of
    the measurement please see \litref{Waechter2008a}.}
  \label{fig:intermittency} 
\end{figure}
In contrast to the amazing progress in design and manufacturing of
WECs, the understanding of the physics of the energy resource, i.e.{}
the turbulent atmospheric boundary layer, and its interaction with the
WECs leaves many open questions and remains insufficient.
To give one example, in the context of wind energy commonly Gaussian
statistics of fluctuations of wind fields are assumed, which is also
reflected in the current industry standard IEC 61400 \cite{IEC2005a}.
It is well-known that in turbulence highly intermittent (i.e.\
non-Gaussian) statistics are found, especially for probability density
functions (PDFs) of velocity increments 
\begin{equation}
\delta u_\tau(t)=u(t+\tau)-u(t)\,. \label{eq:def_incr}
\end{equation}
This effect is even stronger in atmospheric
flows \cite{Ragwitz2001, Boettcher2003, Boettcher2007, Vindel2008,
  Liu2009, Muzy2009, Stewart1970, Tennekes1973}.
Figure~\ref{fig:intermittency} shows that for a $7\sigma$ event of
$\delta u_\tau$ the Gaussian PDF underestimates the probability by
a factor of $10^7$, compared to the measured statistics.
This means if the Gaussian statistics predicts the event to occur
every 1250 years, in reality it is expected to happen once every hour.

Wind velocity increments as a function of the time scale $\tau$ are
chosen here because we are interested in the dynamical response of
wind energy systems to turbulent excitation by atmospheric flows.
Based on a rough estimation using Taylor’s hypothesis of frozen
turbulence (1\,s corrsponds to about 10--20\, m) it is evident that
the
aerodynamic forces typically show dynamics on time scales even below
1\,s. Mechanical, electrical, and electronical components of a WEC
present dynamic behavior with time constants which range up to 1\,min
due to, e.g., yaw control.
From a mathematical point of view the PDFs $p(\delta u_\tau)$
represent two-point statistics including its higher order moments, as
will be explained in \secref{sec:analysis}. These are of particular
relevance for the nonlinear dynamics of a WEC. 

The focus of this paper is twofold. First, we aim at a deeper
understanding of key features of atmospheric turbulence from the
perspective of a WEC system, i.e., on temporal scales up to 1\,min.
We present results on how the wind resource can be properly
characterized, modeled, and experimentally reproduced on these scales. 
Second, we are interested in the effect of these turbulent features on
WEC systems, and how their properties are reflected in the generated
loads and electrical power output.
For the investigation of both aspects, we have developed methods which
are presented in this paper. The aim is to show in which aspects
advanced turbulence research can contribute essentially to wind energy
research. 
We claim that an improved understanding of turbulence is inevitable for
new insights in the response dynamics of WEC systems,
including loads, extreme values, and power output variation. 
Rather than giving a comprehensive overview of existing methods, we
exemplarily present a number of approaches which are focused on the
mentioned intermittent features of atmospheric turbulence.
The point of our work is not claiming to present the best methods of
turbulence investigations, but to show the necessity of further
advanced turbulence research for wind energy systems.

The structure of the paper is as follows.
Section \ref{sec:intermittency} is investigating the typical
small-scale intermittency of atmospheric wind fields. 
Methods for the proper analysis and characterization are presented in
\secref{sec:analysis}, especially focusing on the connection between
homogeneous isotropic turbulence (HIT) and atmospheric turbulence.
In \secref{sec:experiments} we show methods for the proper
experimental reproduction of atmospheric flow conditions in a wind
tunnel setup. In the second part of this section we discuss the impact
of turbulent inflow conditions on a simple experimental dynamic
response system. Results are discussed on the background of HIT
features in atmosperic turbulence.
The impact of atmospheric turbulence on WEC systems is investigated in
\secref{sec:loads}, showing that the typical intermittency features
affect all components of a WEC system as well as forces, loads, and
power output.
Section~\ref{sec:modeling} is devoted to the problem of how to
quantify the impact of atmospheric turbulence on aspects of the wind
energy system. We treat  the
dynamics of power conversion by WECs as a result of the interaction
between machines and atmospheric turbulence in terms of stochastic
response systems. 
Section \ref{sec:outlook} finally gives an outlook to future work
towards a multiscale description of atmospheric turbulence in terms of
arbitrary $n$-scale joint probabilities.


\section{Small-Scale Intermittency}
\label{sec:intermittency}

\subsection{Analysis and Characterization}
\label{sec:analysis}

Atmospheric flows constitute the resource or, loosely spoken, the
``fuel'' of wind energy systems. Detailed knowledge of their
properties and behavior is therefore essential for any improvement in
wind energy conversion.

Wind fields are complex structures, fluctuating in time as
well as in space.  
The industry standard IEC 61400 \cite{IEC2005a} defines quantities and
procedures for considering the influence of atmospheric turbulence on
WECs. 
As a practical approach to wind field characterisation, the \emph{ten
  minute mean value} of the horizontal wind speed,
$\bar{u}=\langle u(t) \rangle_{10\,min}$, is used together with the
\emph{standard deviation} $\sigma$ with respect to the same time
interval. These quantities are one-point statistics and thus can not
reflect the succession of velocity values. 
Moreover, the dynamics of current WECs is not grasped by ten-minute
statistics. Fast load changes on the rotor, the dynamical respons of
the WEC, and resulting power output dynamics rather take place at a
time scale of the order of 1\,s.

The IEC standard proposes for a more detailed characterization of wind
fields universal power spectral densities of the horizontal wind
speed. These are also commonly used for the synthetic generation of
wind inflow data \cite{Nielsen2003}. An important limitation of these
spectral models are the purely Gaussian statistics of the resulting
wind fields, especially in the wind speed increments, which does not
reflect experimental results, as shown in \figref{fig:intermittency}.

While necessity and usefulness of standards are beyond doubt,
in the recent years growing demand for a more comprehensive and more
detailed characterization has become evident. 
To characterize \emph{variations} of $u$ in a general way we consider
wind speed \emph{increments} 
\begin{equation}
  \delta u_\tau(t) = u(t+\tau) - u(t)
  \label{eq:increments}
\end{equation}
as already introduced in \secref{sec:intro}.
The moments $\langle (\delta u_\tau)^k\rangle$ of these increments and
their dependence on the scale $\tau$ are
also called \emph{structure functions} and reflect the variation
statistics of order $k$. It is straightforward to see that the
second-order structure function $\langle(\delta u_\tau)^2\rangle$ is
directly related to the auto-correlation function and thus also to the
power spectral density.
As  $\langle(\delta u_\tau)^2\rangle$ grasps only the width of the PDF
$p(\delta u_\tau)$ it becomes clear that, taking only the power spectra
into account, Gaussian statistics of $p(\delta u_\tau)$ are implied.
Alternatively to the moments, also the $\tau$-dependence of the PDFs
$p(\delta u_\tau)$ can be investigated, which consequently reflect the
variation statistics of all orders.

For practical applications, the high complexity of atmospheric flows
has to be further disentangled and a set of crucial parameters and
methods has to be defined. A number of advances could already be
achieved in this direction.
In \cite{Boettcher2003, Boettcher2007} a model for atmospheric wind
speed increment PDFs was presented, reproducing characteristic properties
of different locations, such as offshore sites or onshore sites in
different terrain.
Based on earlier work of Castaing\cite{Castaing1990} the PDFs are
considered as multiple superpositions of Gaussian distributions,
combining the intermittency at a given mean wind speed and the
typical Weibull distribution of 10\,min mean wind speeds.
An application example for an offshore site located in the Baltic Sea
is presented in \figref{fig:Boettcher}. Especially the intermittency
and the evolution of the shape of the PDFs with the scale $\tau$ is
well reproduced by the model. Nevertheless, slight deviations remain. 
For details, including the derivation of the parameters,
please refer to \litref{Boettcher2007}.
This work is currently continued using new findings from extended data
analysis of other offshore measurement campaings.
\setlength{\breite}{60mm}
\begin{figure}
  \centering
  \hspace*{\fill}%
  \includegraphics[width=\breite]{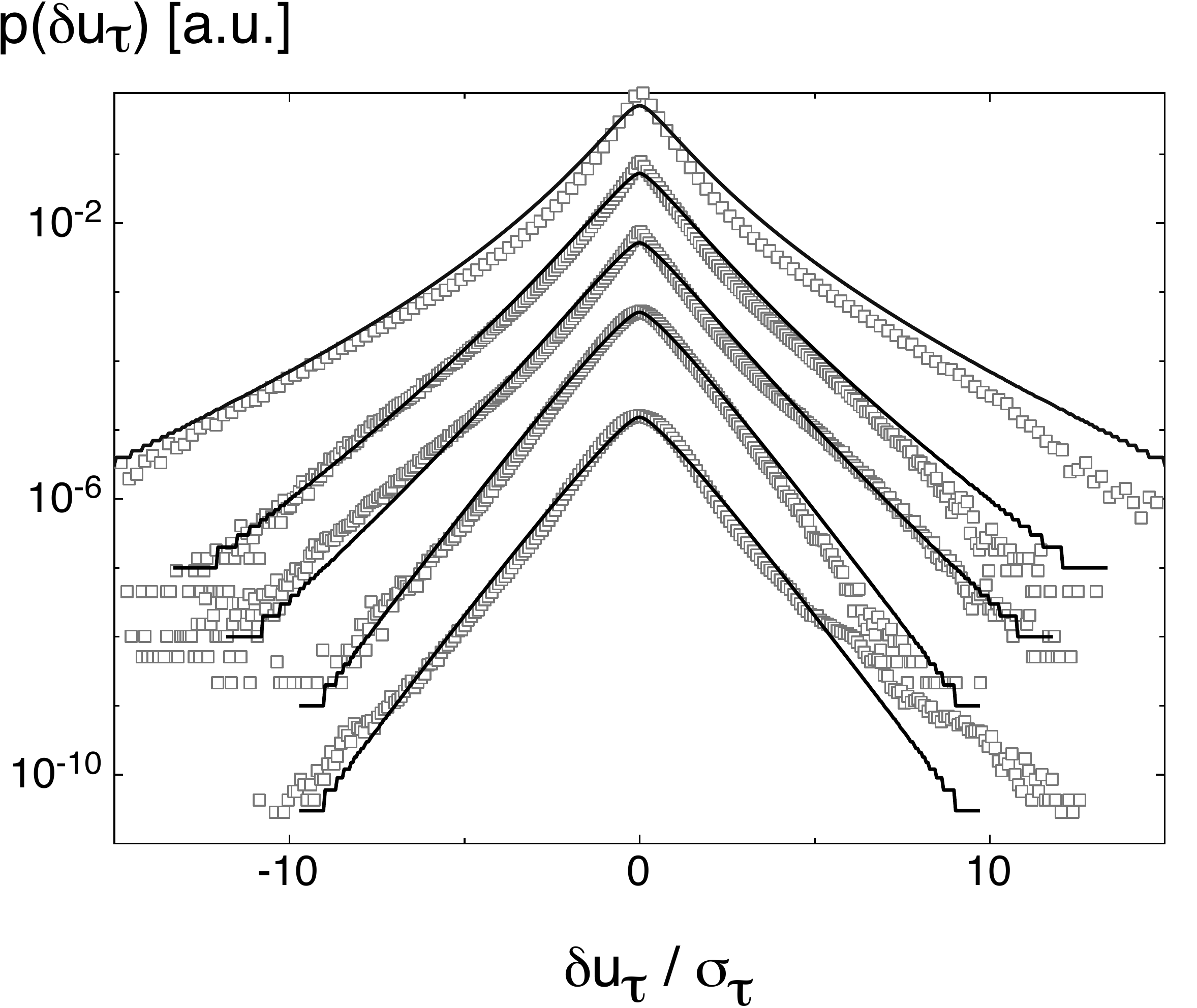}
  \hspace*{\fill}%
  \caption{Modeling of atmospheric probability density functions for
    an offshore site,
    reproduced after \cite{Boettcher2007}.
    Symbols represent the normalized PDFs of atmospheric data
    sets in semi-logarithmic presentation. Solid lines correspond
    to a model according to Eq.~(20) of \cite{Boettcher2007}.
    All graphs are vertically shifted against each other for clarity
    of presentation.
    From top to bottom $\tau$ takes the values of
    (0.2, 10, 20, 200, 2000)\,s.
  }
  \label{fig:Boettcher}
\end{figure}
In a further step, this concept was embedded into a hierarchical
framework for the statistical characterization of atmospheric
turbulence in \litref{Morales2010a}. 
An advanced procedure was proposed for the disentangling of
atmospheric wind measurements into subsets which behave similar to
homogeneous, isotropic turbulence.
It could be shown that for the mentioned subsets the dependence of the
intermittency parameter $\lambda^2$ (for a definition please refer to
\cite{Morales2010a}) on the time scale $\tau$ corresponds with
Kolmogorov's refined scaling law for homogeneous isotropic turbulence
of 1962 \cite{Frisch2001},
\begin{equation}
  \label{eq:K62}
  \langle(\delta u_\tau)^n\rangle \propto \tau^{\frac{n}{3} -\mu\frac{n(n-3)}{18}}\,.
\end{equation}
Here, $\mu$ characterizes the evolution of intermittency in the
scaling behavior of structure functions. The dependence 
$\lambda^2(\tau) = \lambda^2_0 -\frac{1}{9}\mu\log\tau$ is covered by
\eqref{eq:K62} for $n=4$, see appendix of \cite{Morales2010a}. For
different atmospheric data sets the value of $\mu=0.26\pm 0.4$
reported in the literature \cite{Frisch2001} could be confirmed.

In this section we present a similar model for PDFs of
atmospheric wind speed fluctuations $u'(t)=u(t)-\bar{u}$ with
$\bar{u}=\langle u(t) \rangle_{10\,\mathrm{min}}$, which is a common
quantity in the wind energy field. As $u'$ constitutes one-point
statistics it should nevertheless be clearly distinguished from the
two-point statistics of increments $\delta u_\tau$.
A first approach based on cup anemometer data was included in
\cite{Morales2010a}.
In our extended analysis here we use one month of ultrasonic data
measured at the offshore research platform FINO I, where the wind
speed $u$ is taken in the direction of $\bar{u}$. Ultrasonic
anemometers offer higher quality data in terms of higher sampling
rates (10\,Hz in our case) than cup anemometers and avoid effects of
rotational inertia.
In contrast to laboratory turbulence, the pdf $p(u')$ is strongly
intermittent, see figure \ref{fig:fluct_pdf}(a).
Even after conditioning the statistics of $u'$ on a certain value of
$\bar{u}$ the intermittency persists, see $p(u'|\bar{u})$ in figure
\ref{fig:fluct_pdf}(b).
Thus, the source of intermittency is not only the non-stationarity of
the mean wind speed.
As shown in \cite{Morales2010a} in fact the statistics of $u'$ seem to
follow a Gaussian distribution within single 10-minute time spans.
Therefore, the effect of the atmosphere can be understood as mixing
these clean Gaussian statistics by the randomization of the ten-minute
standard deviations $\sigma_T$ over periods longer than 10 minutes. 
The distribution $p(\sigma_T)$ can in most cases at least
approximately be described by a log-normal distribution
\cite{Hansen2005}, compare Figs.~\ref{fig:fluct_pdf}(c) and (d).

\setlength{\breite}{0.45\linewidth}
\begin{figure}
  \centering%
  \includegraphics[width=\breite]{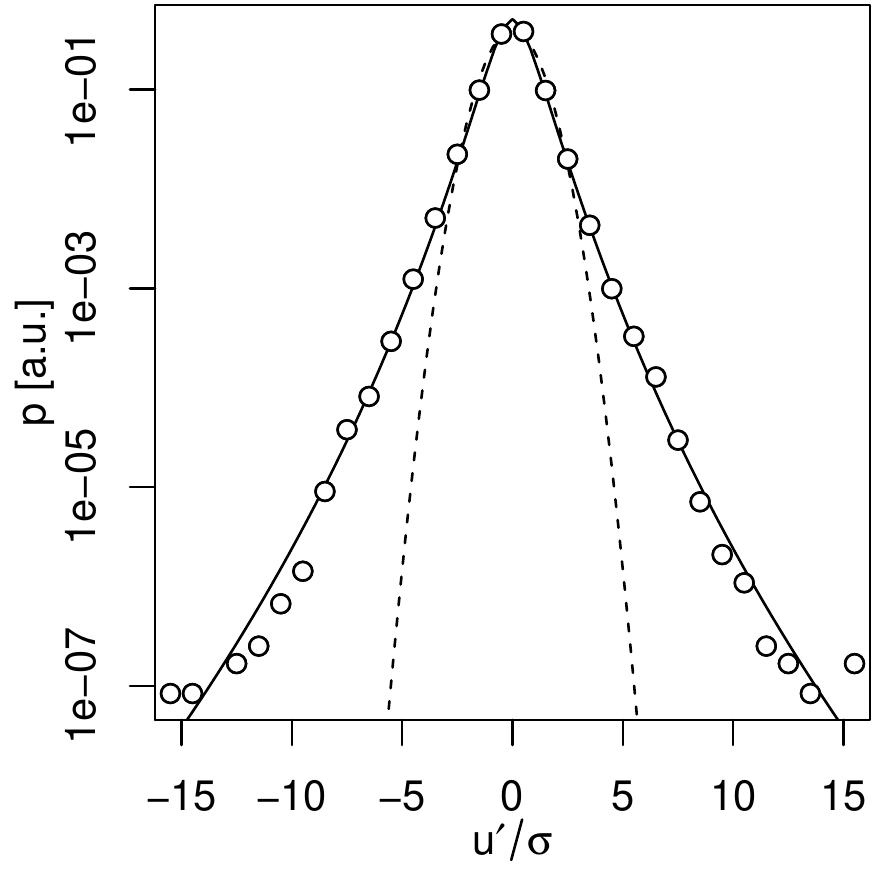}%
  \hspace*{\fill}
  \includegraphics[width=\breite]{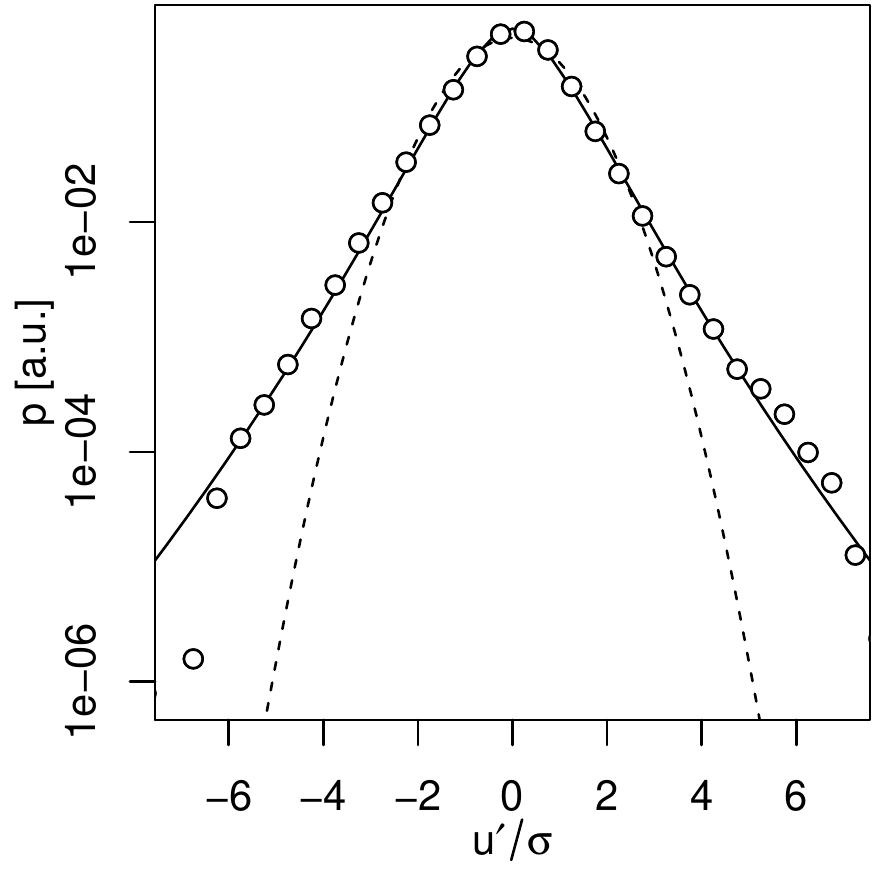}%
  \\
  \includegraphics[width=\breite]{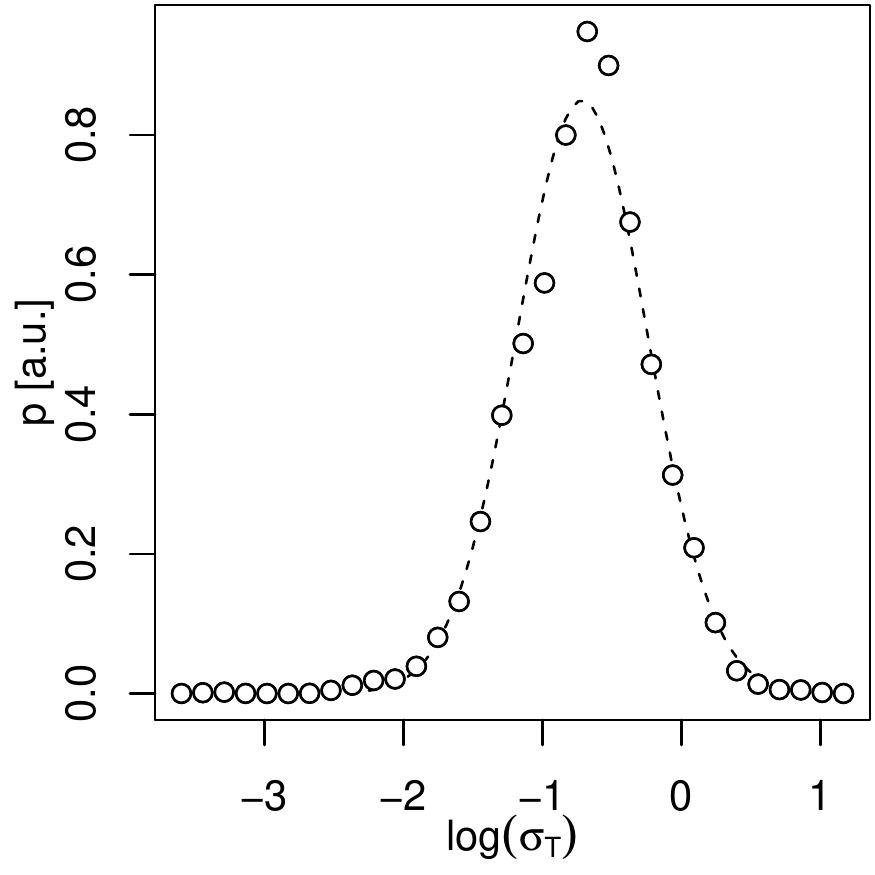}%
  \hspace*{\fill}
  \includegraphics[width=\breite]{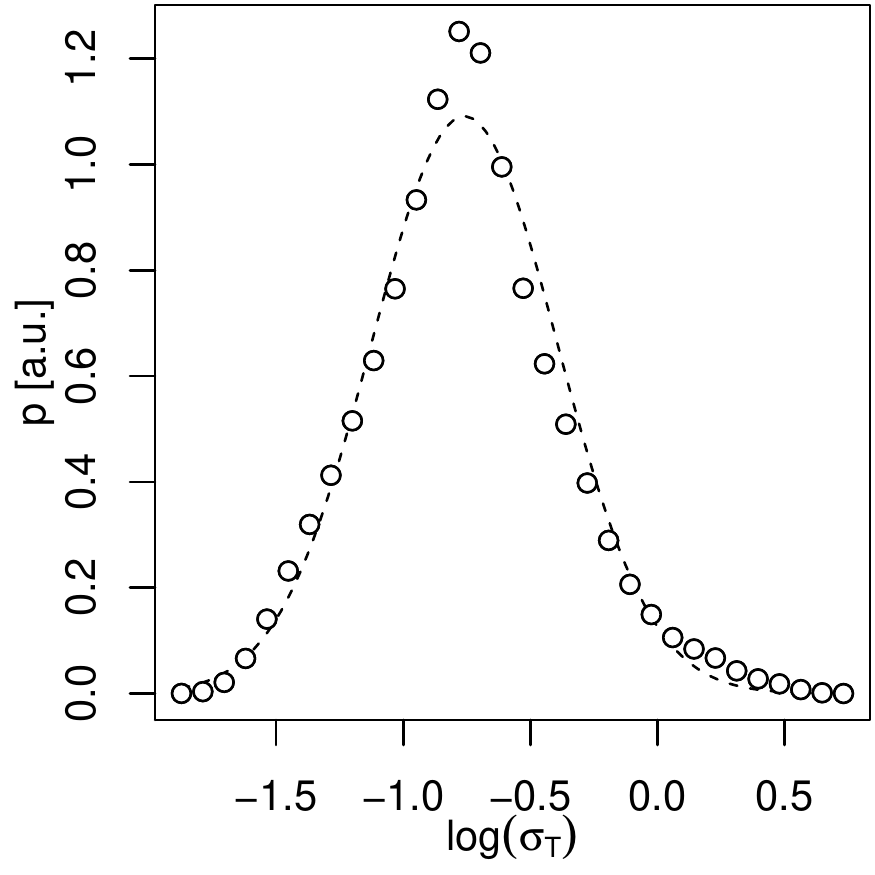}%
  \begin{picture}(0,0)%
    \put(-103,95){(a)}%
    \put( -40,95){(b)}%
    \put(-103,44){(c)}%
    \put( -40,44){(d)}%
  \end{picture}%
  \caption{Statistics of fluctuations $u'$ of atmospheric wind speeds,
    measured at the German offshore research platform FINO I which is
    located in the German Bight about 40\,km north of Borkum island.
    Measurements were taken in 100\,m height, where the yearly average
    wind speed is close to 10\,m/s.
    Parts (a) and (b) present empirical PDFs of fluctuations $u'$
    (symbols), model results according to Eq.~\eqref{eq:PDF_Super}
    (solid lines) and Gaussian PDFs for comparison (dashed lines).
    Part (a) shows $p(u')$ for all data and (b) $p(u'|\bar{u})$
    conditioned on $\bar{u}=(10\pm1)$\,m/s. In parts (c) and (d) we
    show distributions of logarithms of the standard deviations
    $\sigma_T$. Part (c) again represents $p(\log\sigma_T)$ for the
    full data set while (d) shows $p(\log\sigma_T|\bar{u})$
    conditioned as in (b).}
  \label{fig:fluct_pdf}
\end{figure}
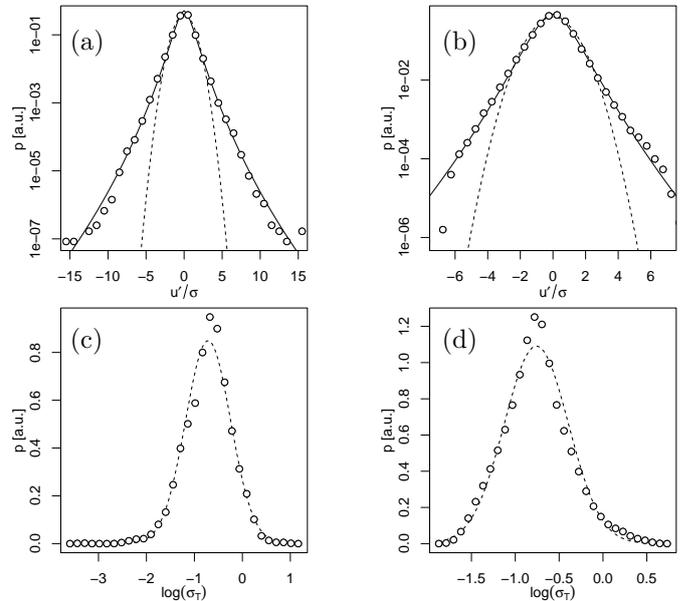

Inspired by the works of Castaing \cite{Castaing1990} and
Boettcher \cite{Boettcher2007} we can now set up an improved model the
PDFs of $u'$ as superpositions of Gaussian distributions with
different standard deviations, where the PDF of these standard
deviations itself is log-normal:
%
\begin{multline}
  p(u') = \\
  \int_{0}^{\infty} \frac{d\sigma_{T}}{\sigma_{T}^2}  \,
  \frac{1}{2\pi \alpha} 
  \underbrace{%
    \exp\left\{-\frac{u'^2}{2\sigma_T}\right\}}_{%
    \parbox{0.2\linewidth}{\footnotesize\centering
      Gaussian PDF\\for single $\sigma_T$}}
  \underbrace{%
    \exp\left\{-\frac{(log\frac{\sigma_T}{\sigma_m})^2}{2\alpha^2}\right\}}_{%
    \parbox{0.2\linewidth}{\footnotesize\centering 
      log-normal PDF\\of $\sigma_T$ values}}\,,
  \label{eq:PDF_Super}
\end{multline}
%
where $\sigma_{m}=\exp\left\{\langle\log\sigma_T\rangle\right\}$ and
$\alpha^2=\langle(\log\sigma_T)^2\rangle-\langle\log{\sigma_T}\rangle^2$.
In this model, $\alpha^2$ plays a similar role as $\lambda^2$ in the
statistics of increments, and quantifies the amount of intermittency in
the PDF.
This compact model reproduces the empirical values quite well, as
presented in figures \ref{fig:fluct_pdf}(a) and (b) for the
unconditioned and conditioned PDFs of atmospheric wind speed
fluctuations, respectively.
As all parameters of \eqref{eq:PDF_Super}
can be derived from the basic 10-minute statistics which are commonly
measured during field campaigns for wind energy applications, our model
gives access to the high-frequency statistics of $u'$ even if no
high-frequency data have been recorded.
The wind speed fluctuations $u'(t)=u(t)-\bar{u}$ as discussed here
have to be clearly distinguished from increments $\delta u_\tau$, see
(\ref{eq:increments}) and \figref{fig:Boettcher}.
Considering together the results for increments shown in
\figref{fig:Boettcher} and fluctuations in \figref{fig:fluct_pdf}, now
models of both the one-point and two-point statistics for wind data
exist. Their validity was shown using examples of offshore data. Both
the wind speed fluctuations $u'$ and increments $\delta u_\tau$ are
relevant for WECs. Using these models, complete and straightforward
characterizations of both quantities are possible. Besides the models
described here, there exist other possibilities, such as special
parametric distributions, but a systematic comparison is beyond the
scope of this paper. We believe that a strength of our approach is its
motivation by the physics of turbulence and in the easy access to the
necessary parameters.

A proper characterization forms the basis for an improved
understanding of atmospheric turbulence, which in term leads to
improved experiments, models, synthetic wind fields, and load
estimations, to name a few examples.


\subsection{Experimental Reproduction}
\label{sec:experiments}

Besides an improved characterization and description of atmospheric
wind fields, it is of special relevance for experimental
investigations to properly reproduce these statistical features in the
wind tunnel. This includes in particular a rescaling of spatial
correlations (e.g., gusts) to dimensions of the wind tunnel and the
models in use, respectively. 
Up to today there exists no method for the generation of turbulence
that allows for a complete control of spatial correlations nor
statistical properties of the resulting flow on all scales due to the
natural decay of laboratory turbulence.
The general aim of our setup is to generate and to reproduce
intermittency features in a controlled way and in a defined response
time window. To the best of our knowledge, this has not been addressed
before by turbulence generating setups.
Classically, atmospheric wind tunnels are used for the generation of
atmospheric like wind fields, where the flow is evolving over passive
roughness elements placed over several meters in the wind tunnel. This
method is very limited when it comes down to the generation of wind
fields with defined intermittency. Furthermore, several approaches
exist which combine different turbulence generating methods, e.g.{}
\cite{Cekli2010, Kang2010, Cal2010}.
This subsection describes current
results concerning options and limits of artificially created
turbulence by an active grid with respect to spatial structures and
reproducibility.

\subsubsection*{Active grid}
\label{sec:active-grid}
\begin{figure}
  \centering
  \includegraphics[width=70mm]{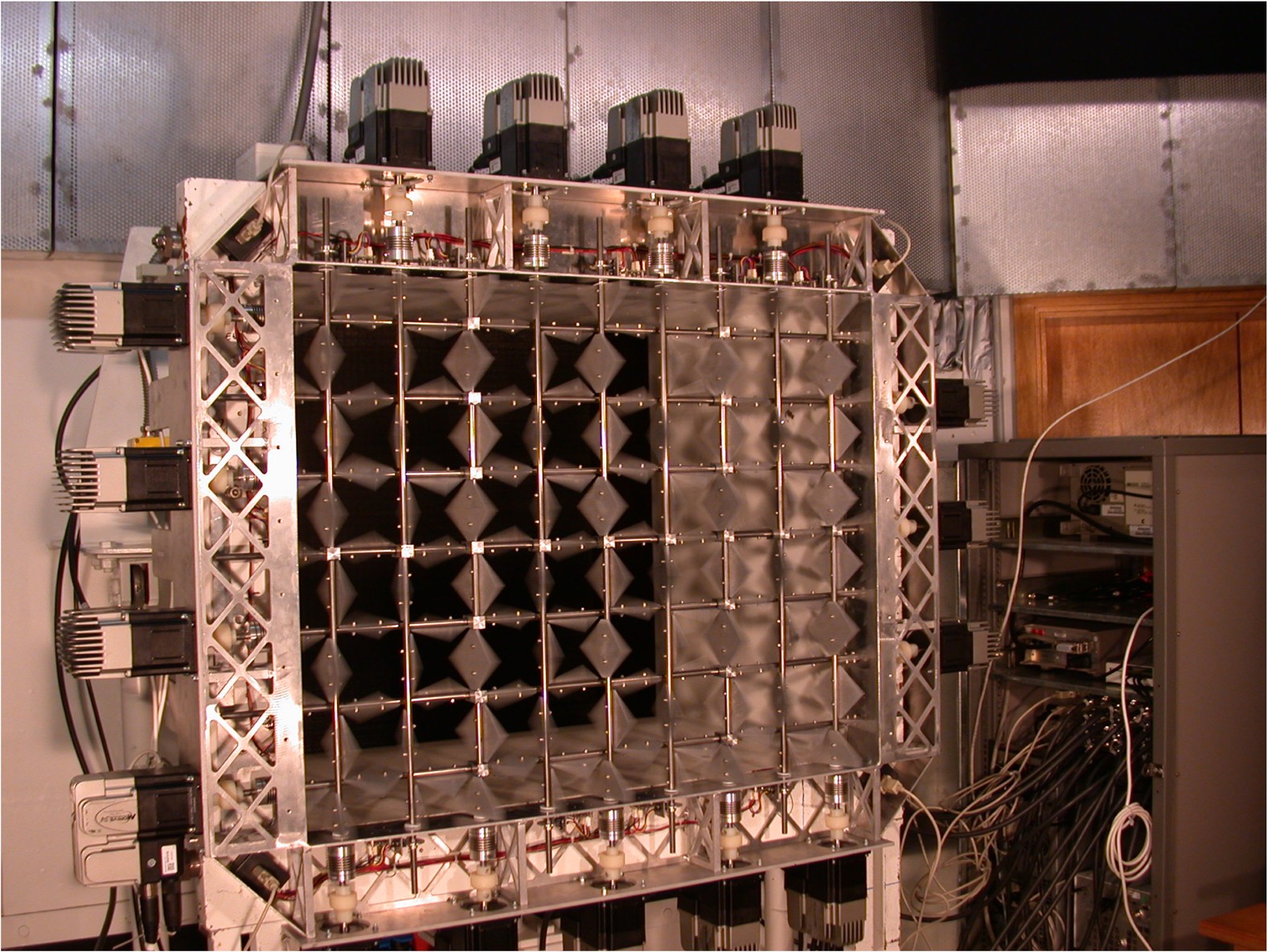}
  \caption{Active grid mounted to the wind tunnel outlet at the
    University of Oldenburg. Visible are the horizontal and vertical
    axes with mounted vanes and individual stepper motors.}
  \label{fig:ActiveGrid} 
\end{figure}

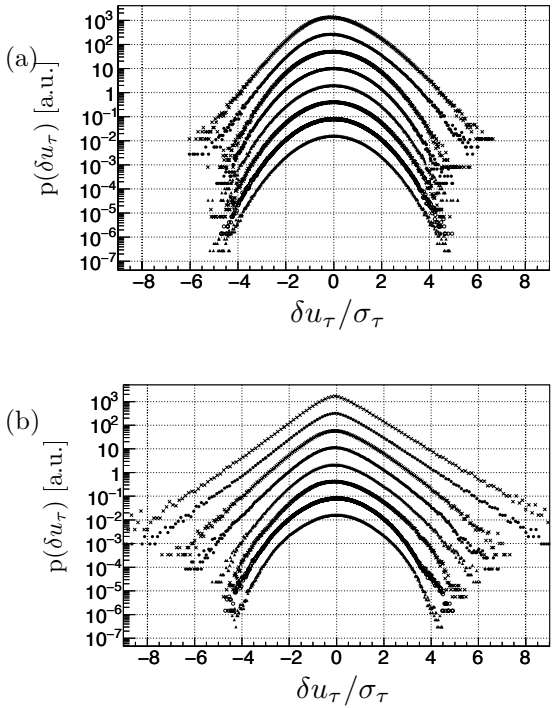
\begin{figure}
  \centering
  \includegraphics[width=70mm]{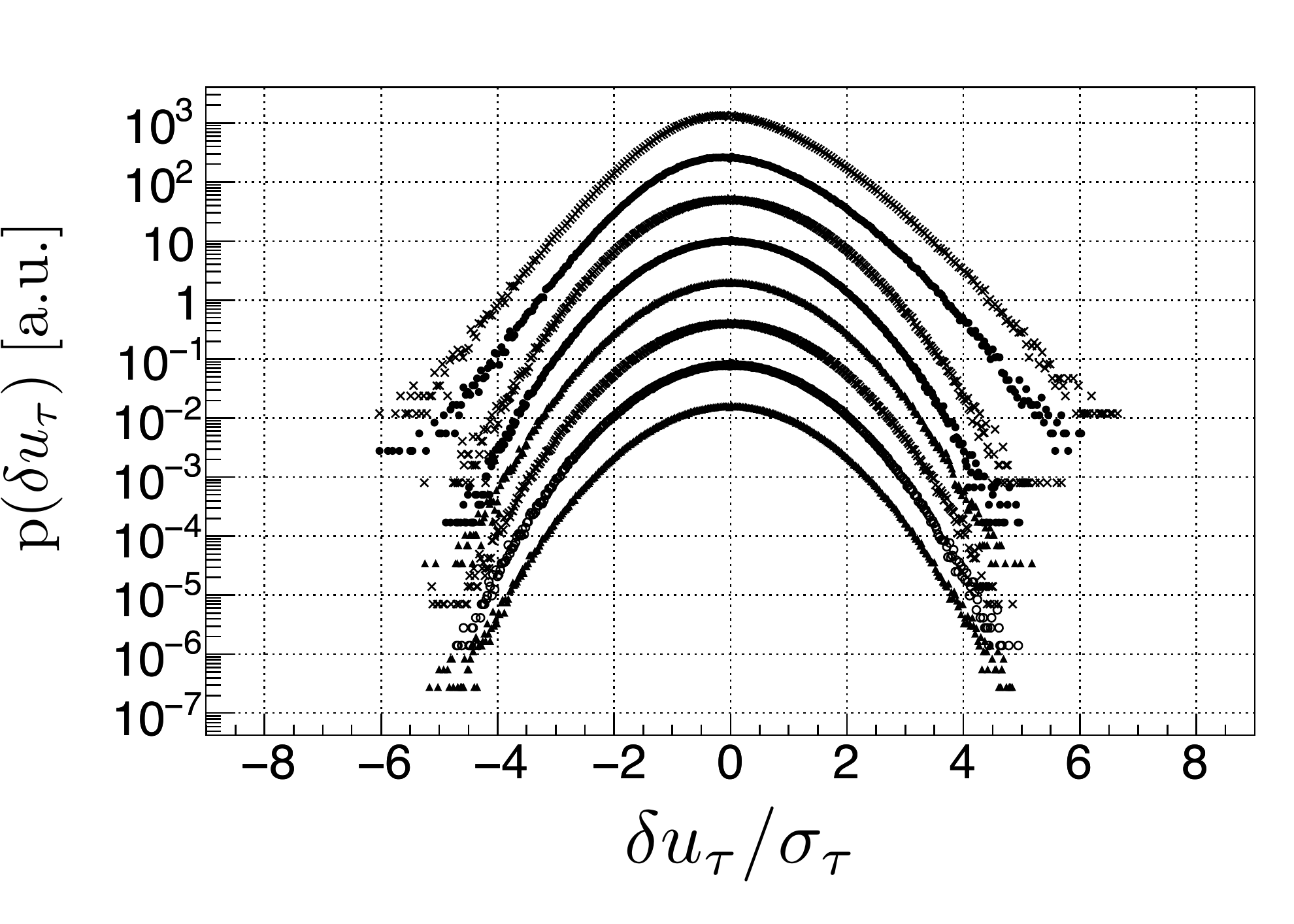}\\
  \includegraphics[width=70mm]{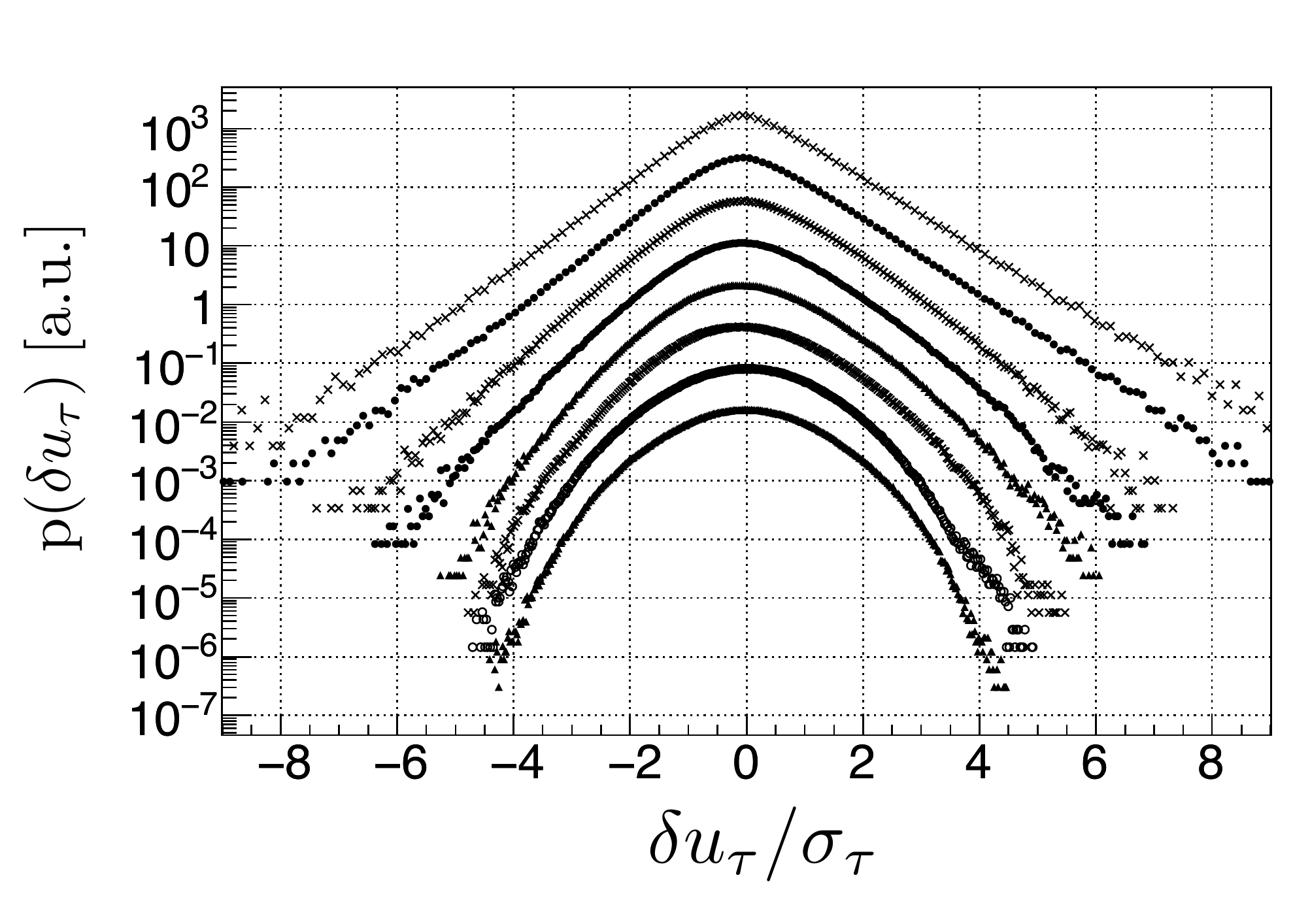}%
  \begin{picture}(0,0)%
    \put(-95,112){(a)}
    \put(-95, 50){(b)}
  \end{picture}%
  \caption{(a) PDFs of velocity increments $\delta u=u(t+\tau)-u(t)$
    for passive mode of the active grid, reproduced after
    \cite{Knebel2011}. Time lags $\tau$ from top to bottom are
    $2.5\cdot 10^{-4}$\,s, $5\cdot 10^{-4}$\,s, $2.5\cdot 10^{-3}$\,s,
    $\ldots, 5\cdot 10^{-1}$\,s. (b) PDFs of velocity increments for
    the active grid with stochastic control of vane movement. Time
    lags $\tau$ are
    as in (a). The graphs are shifted vertically for clarity of
    presentation. 
  }
  \label{fig:PDFActiveGrid}
\end{figure}

To overcome the problem of fixed solidity and fixed effective mesh
size of classical grids for the generation of turbulence, a more
flexible and powerful set-up was built up at the University of
Oldenburg. The so-called active grid, cf.~\cite{Makita1991,
  Warhaft1996}, allows for the generation of turbulent flows by
actively rotating nine vertical and seven horizontal axes with
square-shaped vanes mounted to them (see \figref{fig:ActiveGrid}).
Each of these axes is controlled by a stepper motor and can be
accessed individually with a maximum resolution of $51200$ steps per
rotation.
This grid can be used in passive mode with statically adjusted axes,
as well as in active mode with moving axes. In active mode each axis
can be driven by e.g. sinusoidal or randomized signal resulting in a
time-dependent solidity. This allows for a flexible excitation of the
flow. Figure~\ref{fig:PDFActiveGrid} shows a comparison of the PDFs
for different time lags measured in passive mode with all vanes
oriented parallel to the inflow direction resulting in the minimum
achievable solidity (a) and the PDFs measured in active mode with a
randomized signal (b). It can clearly be seen that in the static
configuration only for the smallest time scales intermittency is
observed. For the active mode, on the other hand, a pronounced
intermittency is generated for a wider range of time scales, as it was
also implemented in the excitation signal \cite{Knebel2011}. This
shift in the turbulent cascade is also reflected in the Reynolds
number, which is $R_{\lambda}=188$ for the passive mode and
$R_{\lambda}=2,243$ for the active mode. The direct coupling of the
motion of the individual axes to the resulting characteristics of the
generated flow field therefore allows for a defined manipulation of
the flow.

\subsubsection*{Reproducibility}
\label{sec:reproduce}
\begin{figure}
  \centering
  \includegraphics[width=68mm]{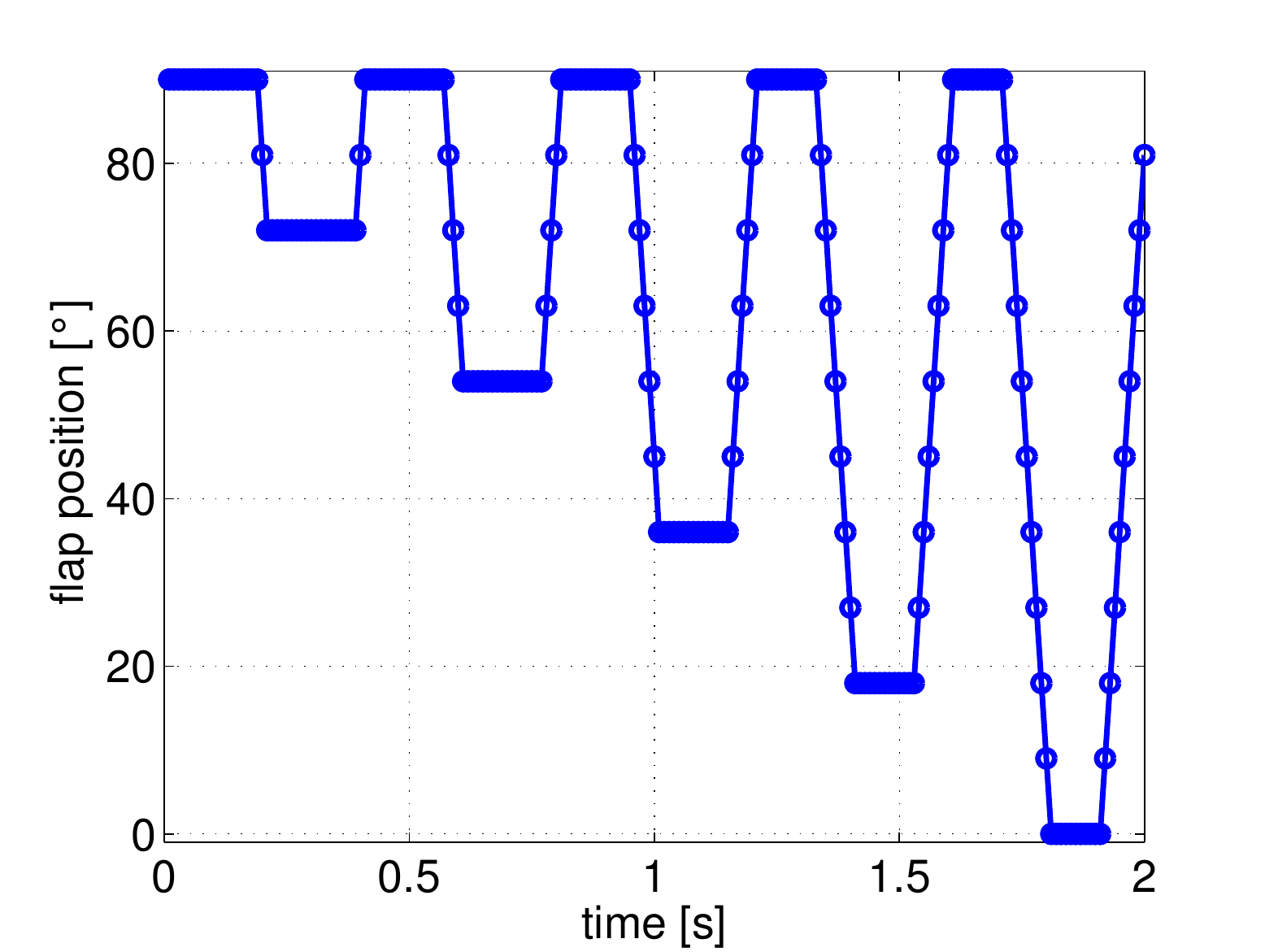}\\
  \includegraphics[width=58mm]{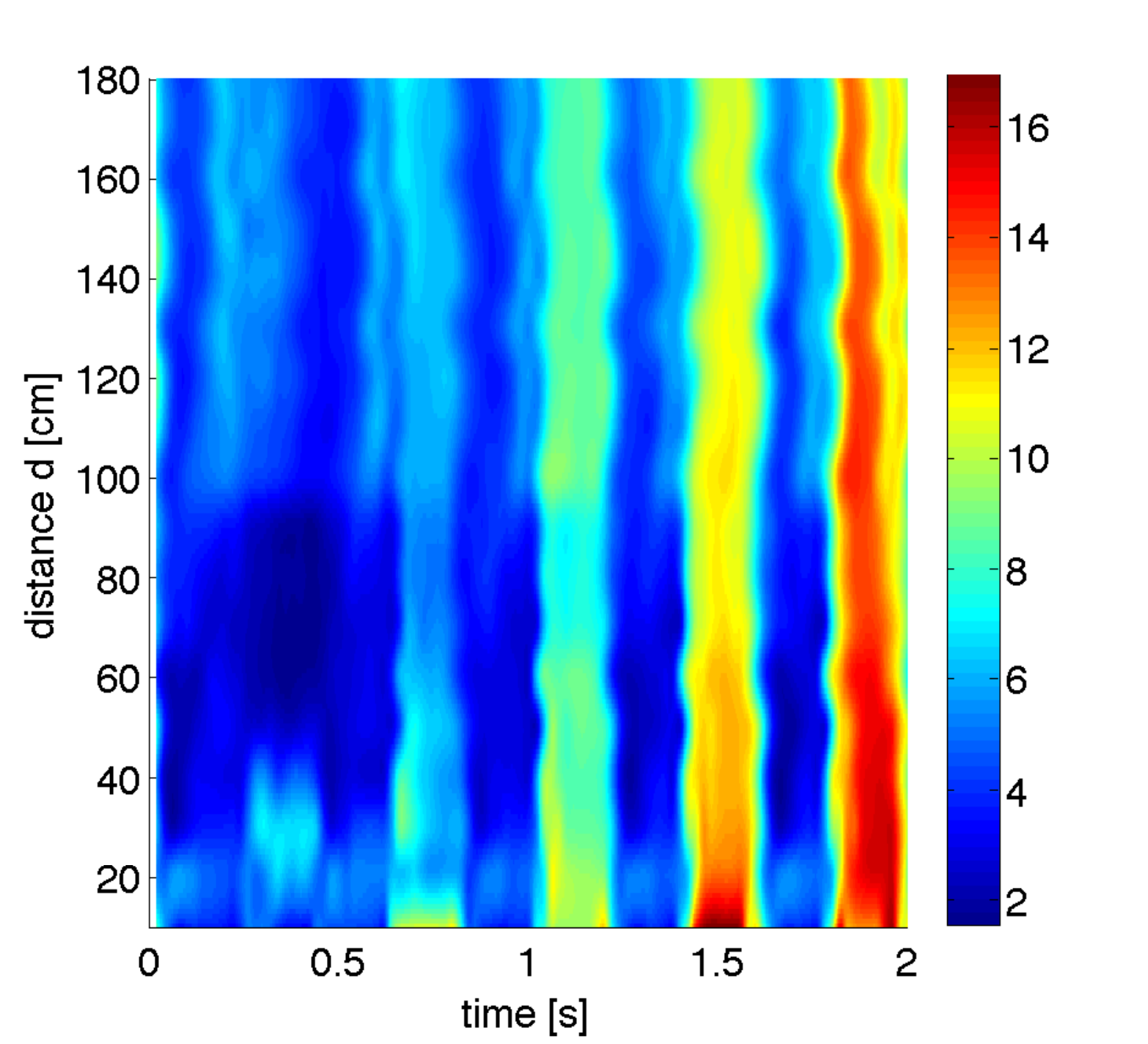}%
  \begin{picture}(0,0)%
    \put(-81,115){(a)}
    \put(-81, 54){(b)}
  \end{picture}%
  \caption{(a) Sequence of realized protocol for the creation of
    increasing velocity pulses. Each pulse has a duration of $0.2$
    seconds. (b) Resulting velocity of the implemented sequence with
    increasing distance to the grid.  
  }
  \label{fig:ActiveGrid_excitation}
\end{figure}
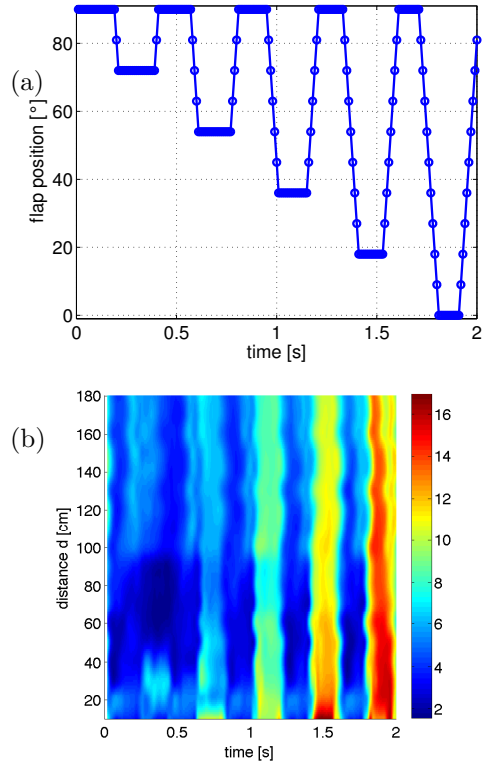

To investigate the reproducibility of turbulent wind fields with
different spatial structures created by the active grid, a control
signal for the axes as shown in
Fig.~\ref{fig:ActiveGrid_excitation}(a) was realized. The complete
protocol consists of several of these sequences in which the angular
position of the axis changes from $90^\circ$ to $72^\circ$,
$54^\circ$, $36^\circ$, $18^\circ$ and $0^\circ$. In the setup
$90^\circ$ corresponds to a vertical orientation of the mounted vanes
with respect to the incoming wind velocity. Here, the time for each
complete pulse was chosen to be $0.2$ seconds. One pulse is defined as
the time from the beginning of a plateau until the end of the
following flank. With a maximum angular velocity of $9^\circ$ per
$10ms$, a pulse length of $0.2$ seconds results in $0.1$ seconds,
where the flaps are in the position of $0^\circ$ and was chosen to be
the shortest reasonable pulse. In order to keep the overall solidity
of the grid constant for the whole sequence the protocol for the outer
eight axes has been programmed with an offset of $90^\circ$. For the
characterization of the resulting flow, velocity measurements on the
centerline behind the grid have been carried out using hot-wire
anemometry with a sampling frequency of $40$\,kHz for different
distances $d$ to the grid varying from $10$\,cm to $180$\,cm with
$\Delta d=10$\,cm and an inflow velocity of $16$\,m/s.
Figure~\ref{fig:ActiveGrid_excitation}(b) shows the resulting velocity
of the implemented sequence with increasing distance to the grid. It
can clearly be seen that the different angles of the axes result in
pulses with increasing velocities. It can also be seen that the
measured flow velocities change with increasing distance to the active
grid, which indicates a decay of the generated velocity pulses and
spatial structures, respectively.

\begin{figure}[t]
  \centering
  \includegraphics[width=70mm]{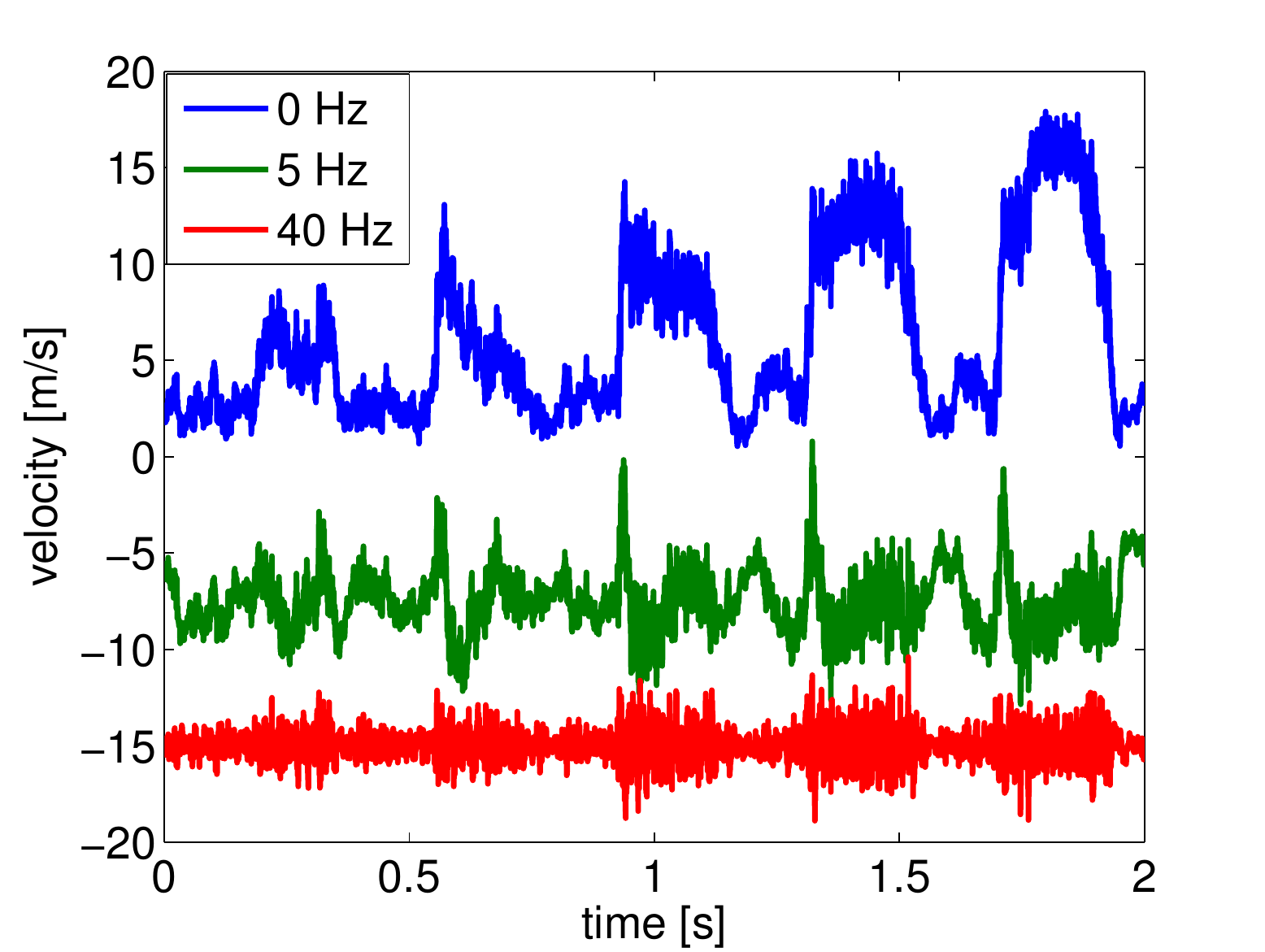}\\
  \includegraphics[width=70mm]{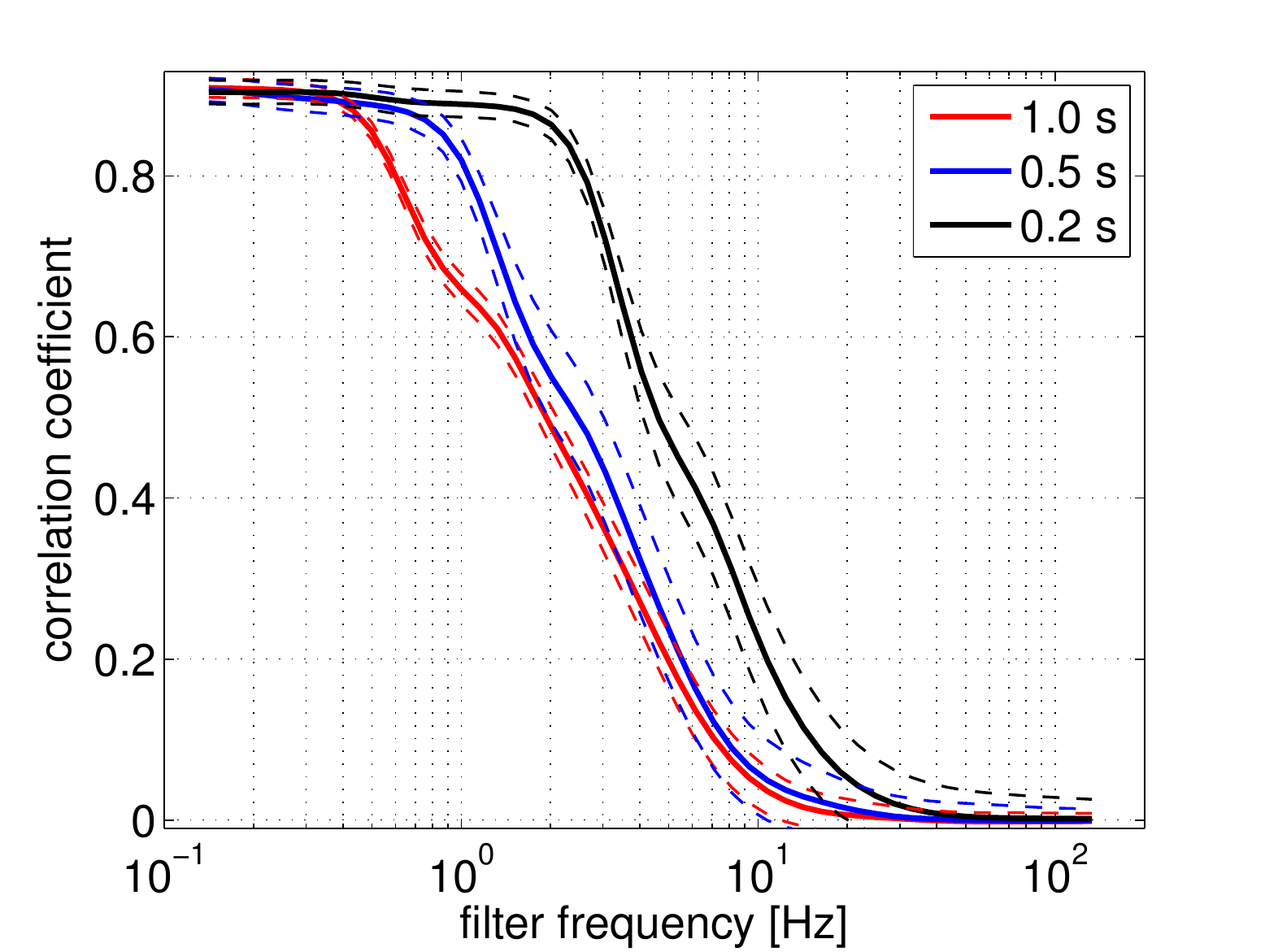}%
  \begin{picture}(0,0)%
    \put(-95,120){(a)}
    \put(-95, 55){(b)}
  \end{picture}%
  \caption{(a) Velocity sequence with pulse duration of $0.2$ seconds
    measured at distance $d=40$\,cm to the grid (blue). The green and
    red curves represent the high-pass filtered signal at filter
    frequencies of $5$\,Hz and $40$\,Hz, respectively. Graphs are
    shifted vertically for clarity of presentation. (b) Averaged
    correlation coefficient over cutoff frequency of high-pass filter
    for excitation sequences with pulse duration of $0.2$\,s (black),
    $0.5$\,s (blue) and $1$\,s (red). The dashed lines indicate the
    corresponding standard deviation.  
  }
  \label{fig:ActiveGrid_sequence}
\end{figure}
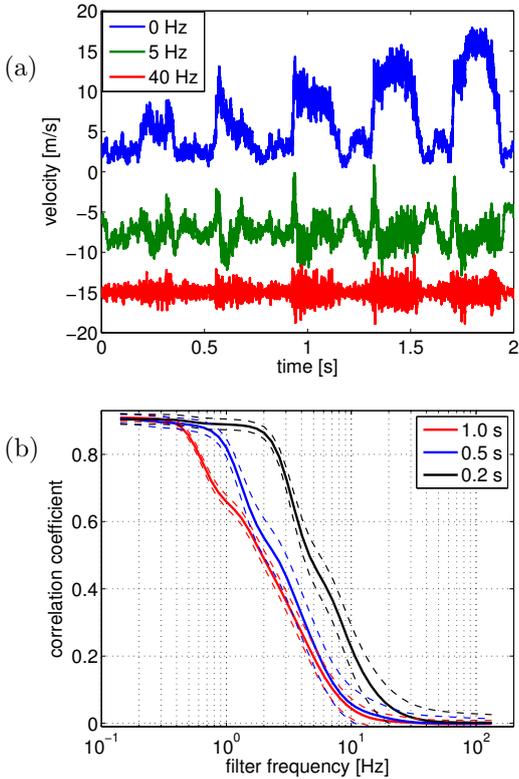

To investigate the decaying process of these structures in more
detail, three excitation protocols with several identical sequences
like presented in \figref{fig:ActiveGrid_excitation}(a) have been
realized where just the length of the pulses has been changed.
Sequences with pulses of $0.2$, $0.5$, and $1$ second duration have
been implemented and used for the excitation of the grid's axes. In
order to determine where the uncorrelated decay of natural turbulence
starts to dominate, each excitation sequence has been identified in
the measured velocity time series and separated. To filter out spatial
structures from the flow, each of these velocity sequences has been
high-pass filtered with different filter frequencies starting from
$0.14$\,Hz up to $133$\,Hz. Figure \ref{fig:ActiveGrid_sequence}(a) shows
an example of measured velocities for one excitation sequence with a
pulse length of $0.2$ seconds at distance $d=40$\,cm together with the
high-pass filtered signal for filter frequencies of $5$\,Hz (green)
and $40$\,Hz (red). The signals are shifted vertically for clarity of
presentation. It can clearly be seen that higher filter frequencies
result in damping of the impressed pulses. 

Furthermore, correlation coefficients have been determined for
different realizations of measured velocities using identical filter
frequencies.
Figure~\ref{fig:ActiveGrid_sequence}(b) shows the average of the
resulting correlation coefficients for sequences of all three pulse
durations plotted over the filter frequency of the applied high-pass
filter. The velocity sequences have been measured at a distance of
$d=40$\,cm to the grid. The dashed lines represent the corresponding
standard deviation for each graph. It can clearly be seen that the
filter frequency at which the correlation coefficient drops off
increases with shorter pulse durations. Applying Taylor's Hypothesis,
this is also in agreement with the equivalent in spatial terms, where
shorter excitation pulses correspond to smaller spatial structures and
therefore need higher filter frequencies to be filtered out.

To investigate the stability of the generated spatial structures with
increasing distance to the grid, the measured velocity sequences have
been split into isolated single pulses. Conditioned on pulses with
identical motion of the vanes, this procedure allows for a similar
analysis as already applied to the whole sequence. Based on the split
sequences a plot according to \figref{fig:ActiveGrid_sequence}(b) can be
calculated for each distance $d$ to the grid, conditioned on pulses $1$,
$2$, $3$, $4$, and $5$, respectively, for all three pulse durations.
The filter frequency for which the correlation coefficient drops off to
$90\%$ of its respective maximum was defined to be the cutoff
frequency. It defines the smallest time, and thus the smallest spatial
structure, of the generated flow that is not dominated by natural
decay and therefore can reproducibly be generated by the active
grid. 
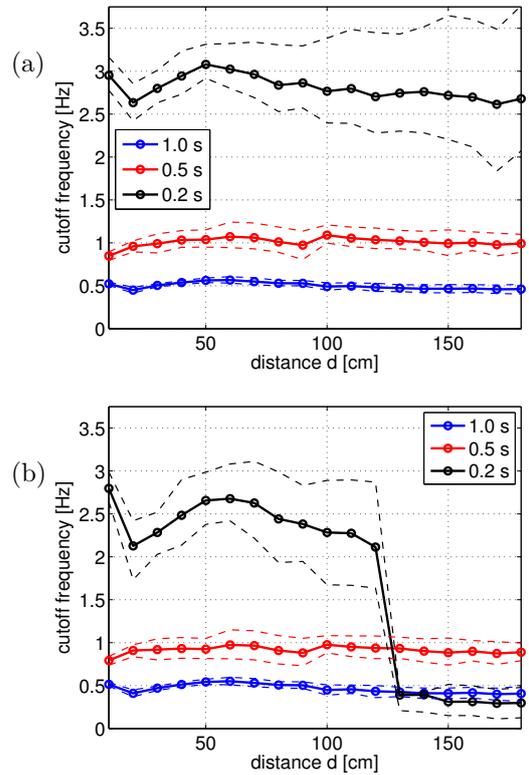
\begin{figure}[t]
  \centering 
  \includegraphics[width=70mm]{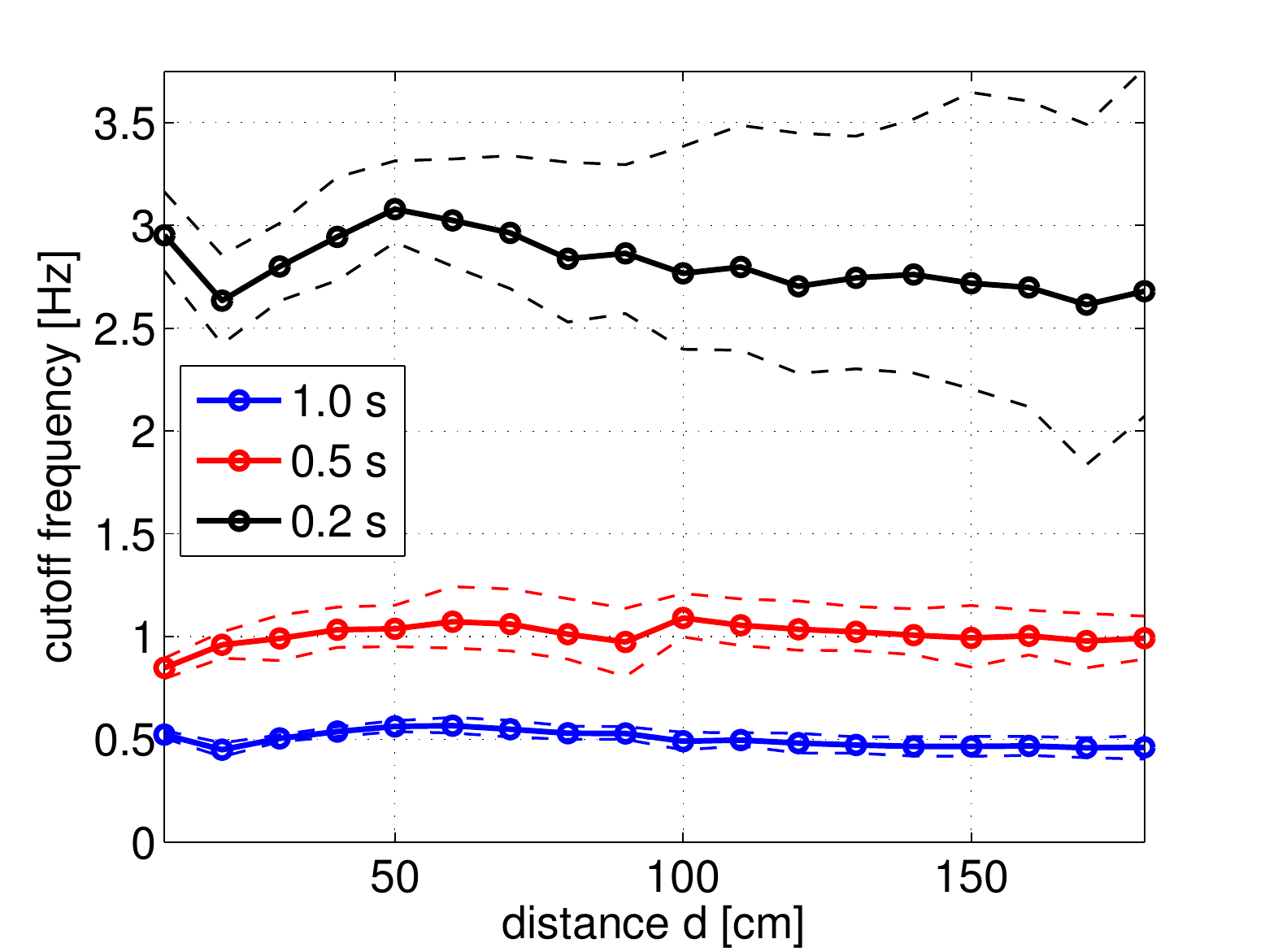}\\
  \includegraphics[width=70mm]{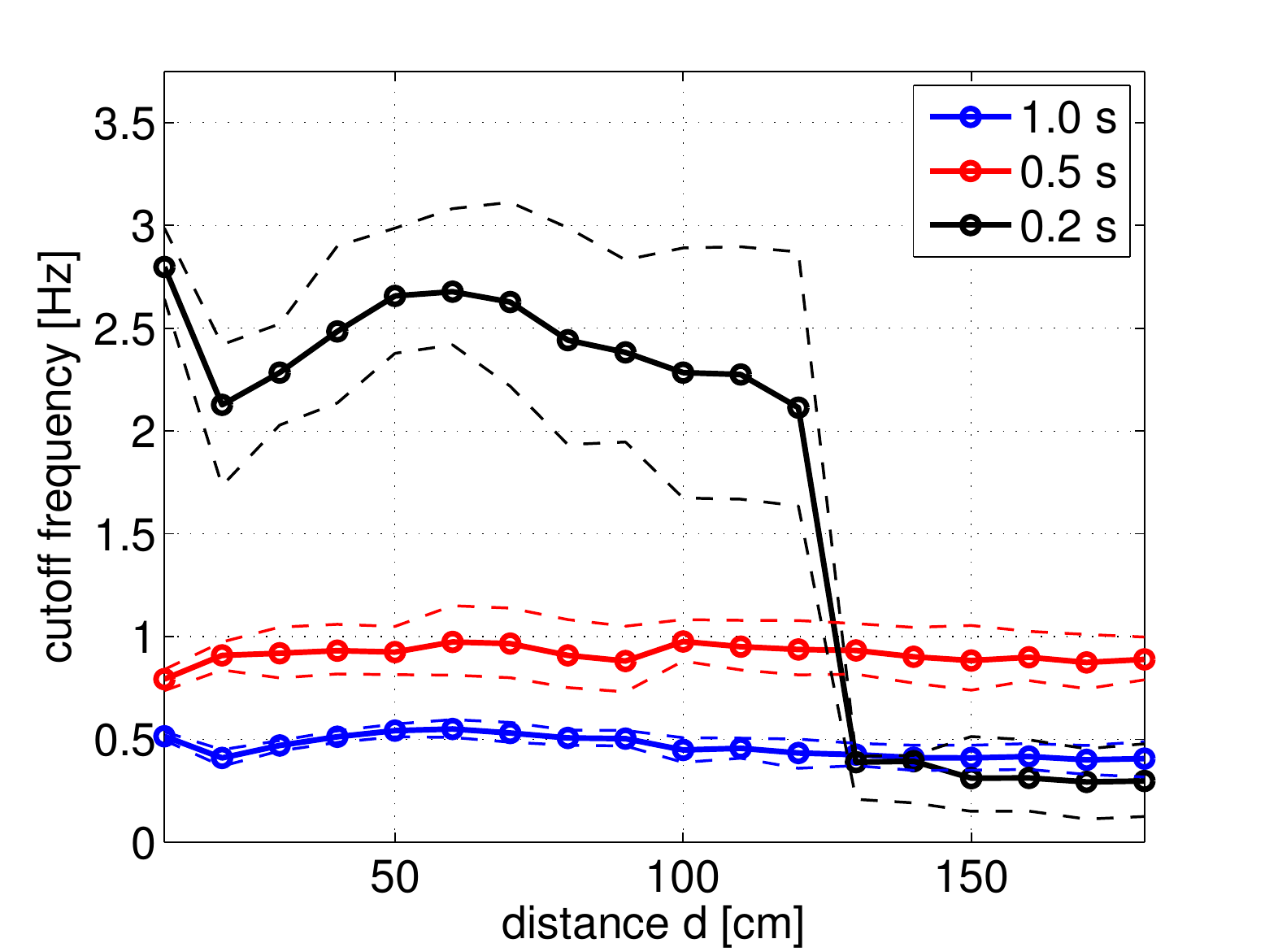}%
  \begin{picture}(0,0)%
    \put(-95,120){(a)}
    \put(-95, 50){(b)}
  \end{picture}%
  \caption{Cutoff frequency over distance $d$ to the grid. The cutoff
    frequency describes the filter frequency of the applied high-pass
    filter, where the averaged correlation coefficient of sequences
    conditioned on (a) the pulse with the highest velocity
    difference and (b) the pulse with the second highest
    velocity difference decreased to $90\%$ of its respective maximum.
    The curves belong to pulses with duration of $0.2$s (black),
    $0.5$s (blue) and $1$s (red). The dashed lines indicate the
    corresponding error. 
  }
  \label{fig:ActiveGrid_correlation}
\end{figure}
The plots in figure \ref{fig:ActiveGrid_correlation} show the cutoff
frequency over the distance $d$ to the grid for (a) conditioned on the
pulse with the largest resulting velocity difference and (b)
conditioned on the pulse with the second largest velocity difference
for all pulse durations. The results show that the cutoff frequency
for the pulse with the largest velocity difference is nearly constant
over the whole distance and the standard deviation for the shortest
pulse duration increases with distance
(Fig.~\ref{fig:ActiveGrid_correlation}(a)). In
Fig.~\ref{fig:ActiveGrid_correlation}(b) a sudden decrease in the
cutoff frequency is visible at a distance $d=130$\,cm for the pulse
with the shortest duration of $0.2$\,s, whereas the cutoff frequencies
for the pulses with longer pulse duration stay constant. This
indicates, that the spatial structure created by the $0.2$ second
pulse is not stable over the whole measurement distance, which has to
be accounted for in the design of excitation protocols.

\subsubsection*{Test of standard anemometers}
\label{sec:test_anemometer}

The ability of the active grid to generate turbulent flow situations
with reproducible temporal and spatial features, as shown in the
previous subsection, provides a powerful tool not only for basic
turbulence investigations, but also for the characterization of
airfoils and model wind turbines exposed to intermittent wind fields.
As an application example, comparative measurements of the same flow
situation have been carried out with two standard anemometers for wind
energy applications: a \emph{Thies First Class Advanced} cup
anemometer and a \emph{Gill Windmaster Pro} 3D ultrasonic anemometer.
The signal recorded with a standard 1D hot-wire probe served as a
reference, since the hot-wire provides the highest spatial and temporal
resolution.

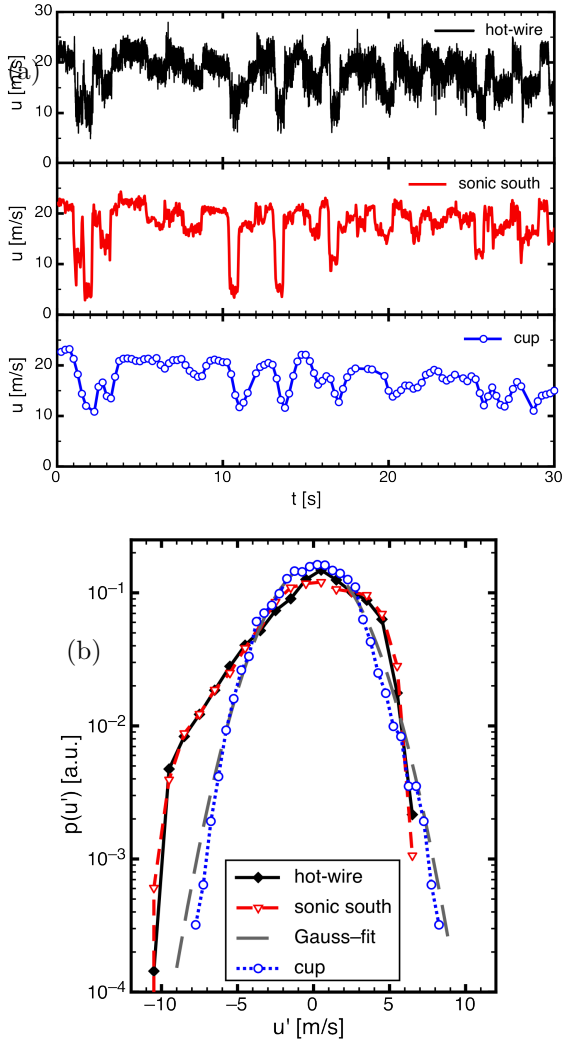
\begin{figure}[t]
  \centering
  \includegraphics[width=75mm]{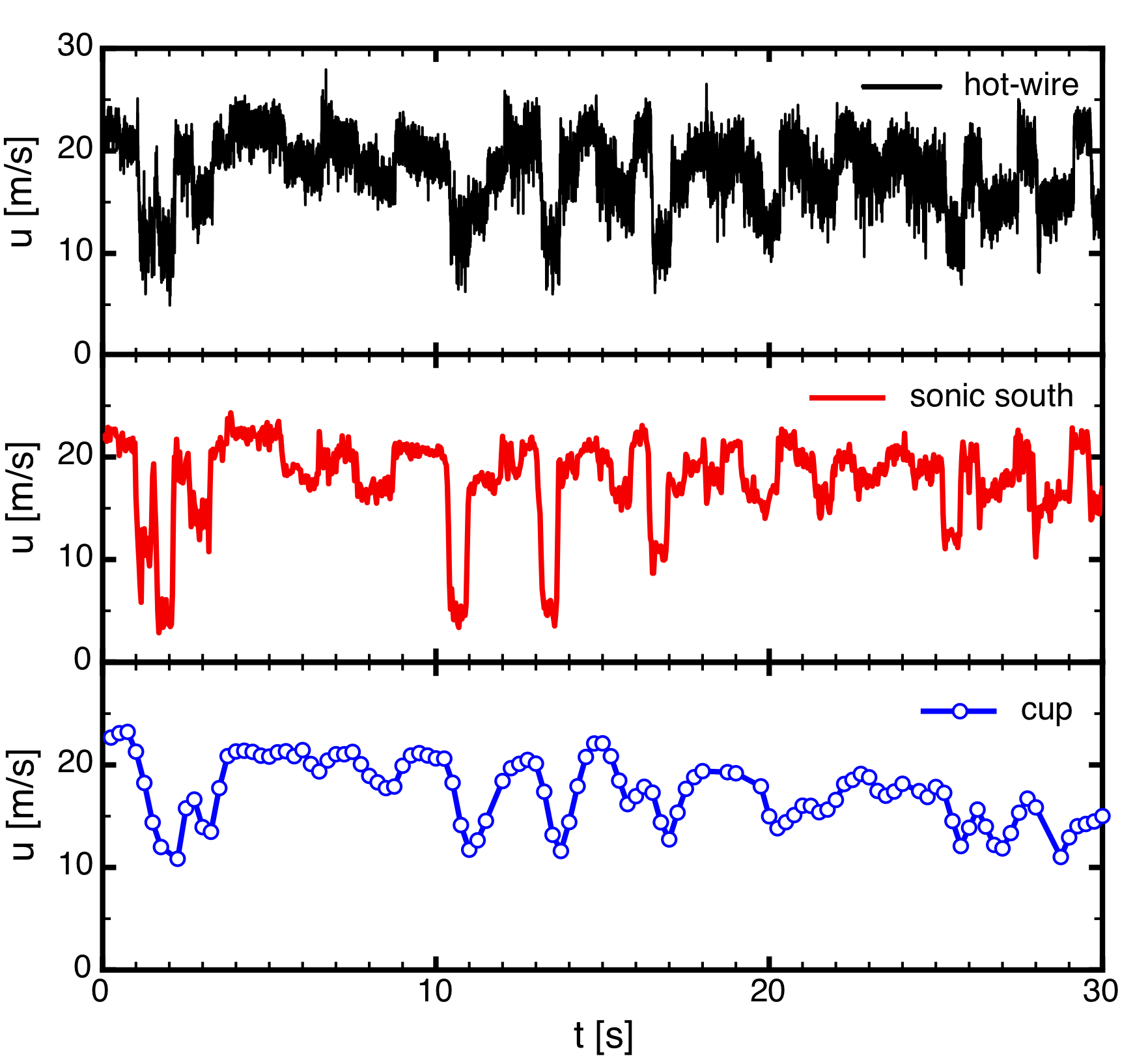}%
  \\
  \hspace*{\fill}%
  \includegraphics[width=61mm]{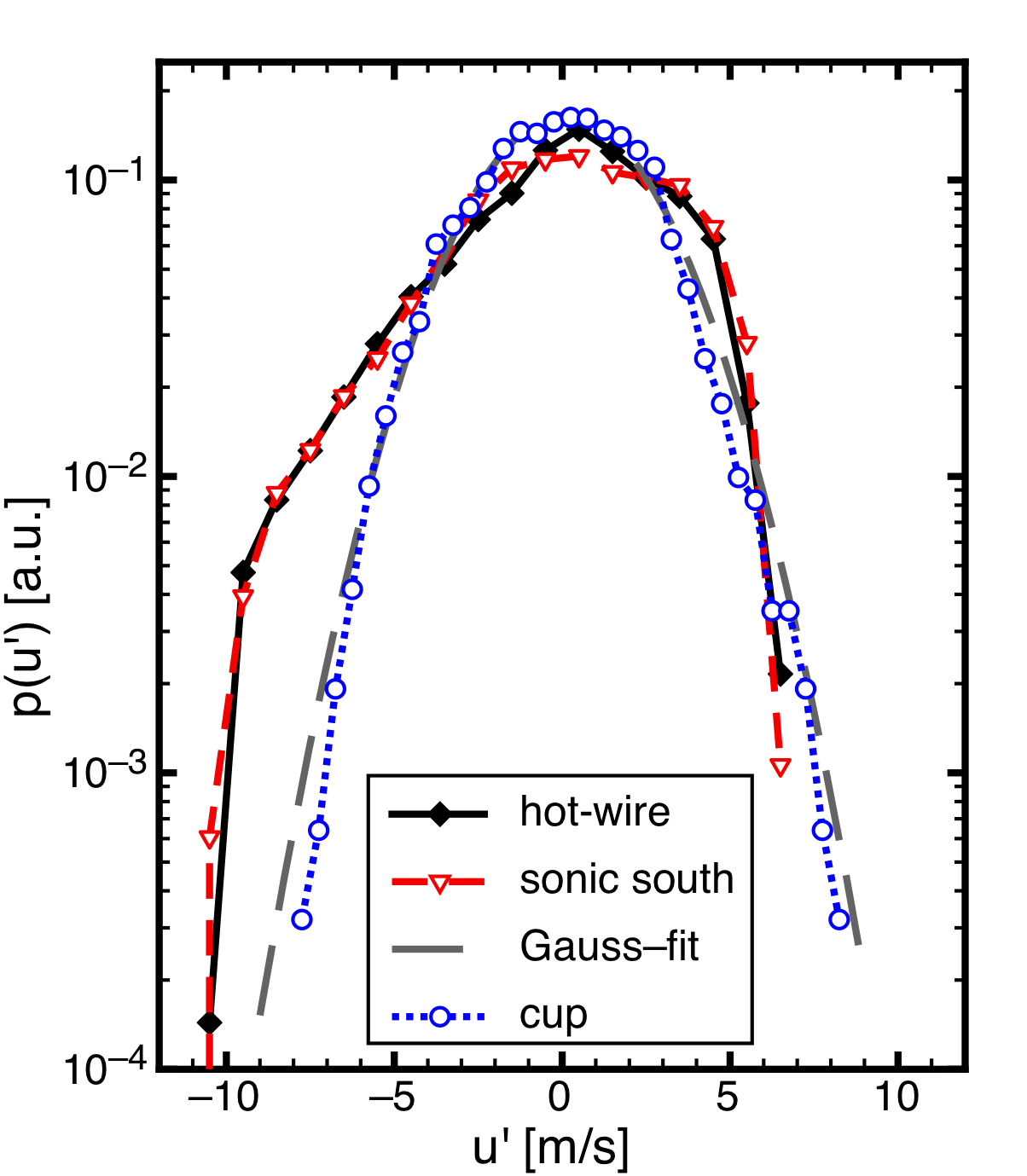}%
  \hspace*{\fill}%
  \begin{picture}(0,0)%
    \put(-105,164){(a)}%
    \put( -95, 65){(b)}%
  \end{picture}
  \caption{(a) 30-seconds-excerpt of the wind speed data measured with
    1D hot-wire anemometer (black), 3D ultrasonic anemometer (red) in
    southward orientation and cup anemometer (blue). The general
    evolution of the turbulent flow is covered by all anemometers, but
    the cup anemometer cannot resolve the fast fluctuations with small
    amplitudes. (b) PDF of the wind speed fluctuations $u'$ for the
    down-sampled (4\,Hz) time series from (a) with Gaussian fit
    (dashed gray).}
  \label{fig:Anemometer_time}
\end{figure}
While 1D hot-wire and cup anemometer lack any directional information,
the data from the 3D ultrasonic anemometer includes the horizontal
wind speed magnitude and direction. However, the alignment of
transducers and supporting structures of the sonic anemometer can
cause flow disturbances under certain orientations \cite{Wiesner2001}.
The used Gill Windmaster Pro anemometer features three support rods
with an angular distance of 120$^\circ$, where the position of the
sensor's north spar coincides with the position of one supporting rod.
Thus, the northward orientation (0$^\circ$) of the anemometer implies
a support structure being upstream the measurement volume, while a
southward (180$^\circ$) orientation implies free inflow to the
measurement volume.

All sensors were calibrated before the measurement and have been
successively placed at a distance $d=$130\,cm downstream of the active
grid with the center of their respective measurement volumes on the
centerline of the wind tunnel test section. The excitation protocol of
the active grid was chosen in such a way, that the resulting modulated
wind flow obeys non-Gaussian statistics as they can be found in
atmospheric wind fields.

For each mean wind speed, 30 minutes of data were recorded with
hot-wire, cup anemometer and ultrasonic anemometer in northward and
southward orientation. The hot-wire data was sampled at 20\,kHz and
smoothened using an average over 20 samples. The lower-resolving cup
anemometer and sonic anemometer were sampled at 4\,Hz and 32\,Hz,
respectively. Figure~\ref{fig:Anemometer_time}(a) shows a 30 second
excerpt of the time series recorded with hot-wire (black), ultrasonic
anemometer (red) in southward orientation and cup anemometer (blue). A
good agreement of the time series for the hot-wire and the sonic
anemometer is observed and the fact that all anemometers capture the
strong fluctuations at e.g.\ $t=$2\,s or $t=$13.5\,s gives evidence
about the robustness and reproducibility of the used excitation
protocol. However, the cup anemometer can only resolve the slower and
more pronounced gust events. The faster fluctuations of the wind speed
(e.g.\ $t=22$\,s to $t=25$\,s) cannot be resolved, which on the one
hand is due to its low temporal resolution and on the other hand is a
result of the spatial extent of the anemometer itself.

A comparison of the basic statistical quantities commonly used in wind
energy, i.e. mean wind speed $\bar{u}$, standard deviation $\sigma_u$
and turbulence intensity $TI= \bar{u}/\sigma_u$, for the time series
shown in Fig.~\ref{fig:Anemometer_time}(a) is given in
Tab.~\ref{tab:Anemometer_time}. Only a slight deviation from the
hot-wire reference is observed for the ultrasonic anemometer's data,
while all quantities of the cup anemometer are significantly lower
than the reference. A closer look at the statistics for all sensors is
provided in the PDFs of the measured wind speed fluctuations $u'$ in
Fig.~\ref{fig:Anemometer_time}(b). The higher resolved hot-wire data
and ultrasonic anemometer data have been down-sampled to the
resolution of the cup anemometer at 4\,Hz, in order to allow for a
comparison of the PDFs. Even the down-sampled hot-wire reference
(black) shows a clearly non-Gaussian PDF with higher probabilities for
strong wind speed decreases, which is also observed in the PDF of the
ultrasonic anemometer (red). In contrast, the PDF of the fluctuations
covered by the cup anemometer (blue) obeys a Gaussian distribution as
fitted in the plot (dashed gray).

\begin{table}[t]
  \centering\small
  \caption{Measured mean wind speed $\bar{u}$, standard deviation
    $\sigma_u$ and turbulence intensity $TI$ for the time series from
    Fig.~\ref{fig:Anemometer_time}(a).}
  \begin{tabular}{|l||c|c|c|} 
    \hline
    sensor & $\bar{u}$ & $\sigma_u$ & $TI$ \\
    \hline
    \hline
    hot-wire & 17.1\,m/s & 3.8\,m/s & 22\,\% \\
    \hline
    sonic south & 17.6\,m/s & 3.5\,m/s& 20\,\% \\
    \hline
    cup & 15.5\,m/s & 2.3\, m/s & 15\,\% \\
    \hline	
  \end{tabular}
  \label{tab:Anemometer_time}
\end{table}

\begin{figure}[t]
  \centering
  \includegraphics[width=68mm]{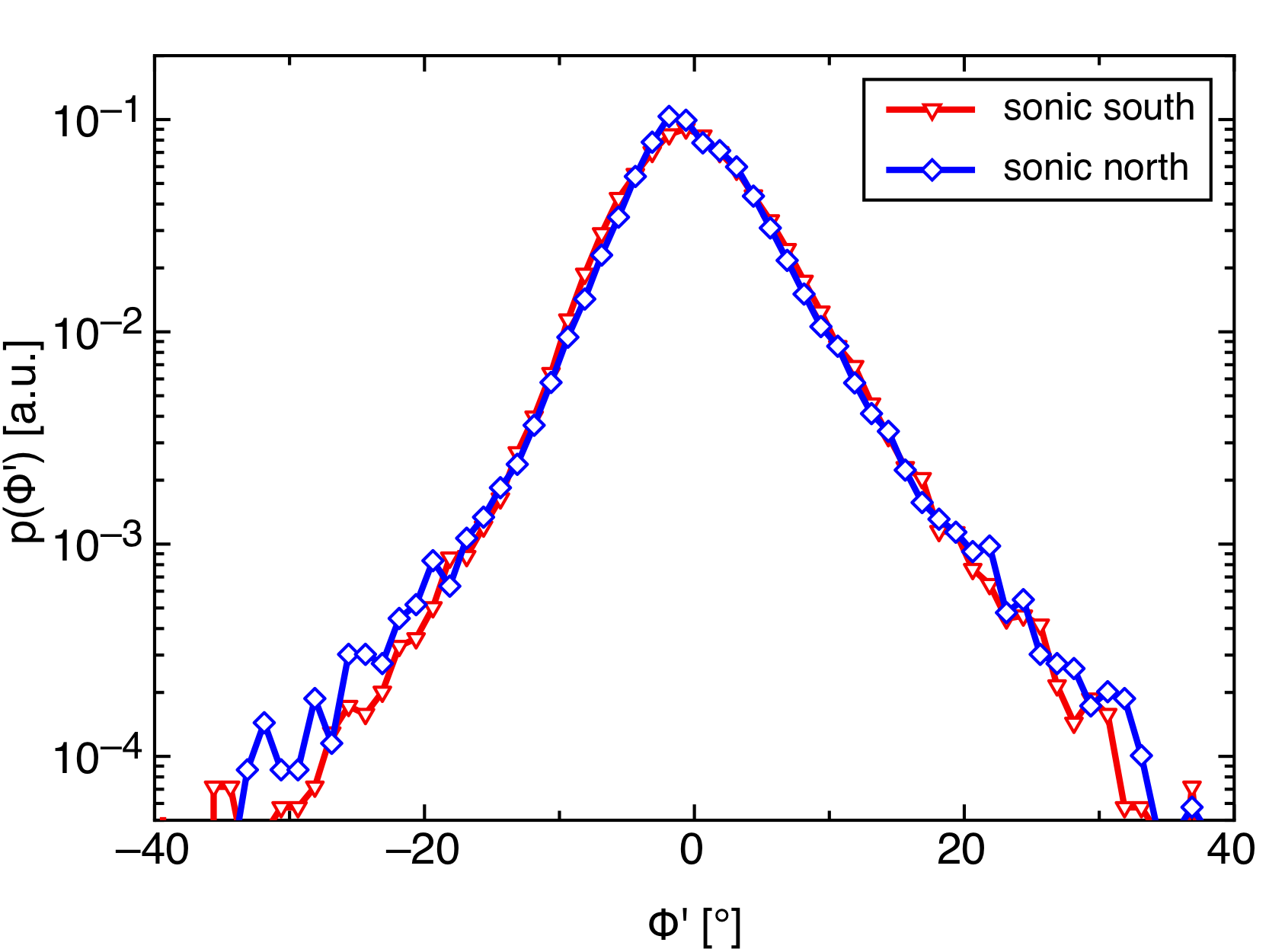}\\
  \includegraphics[width=68mm]{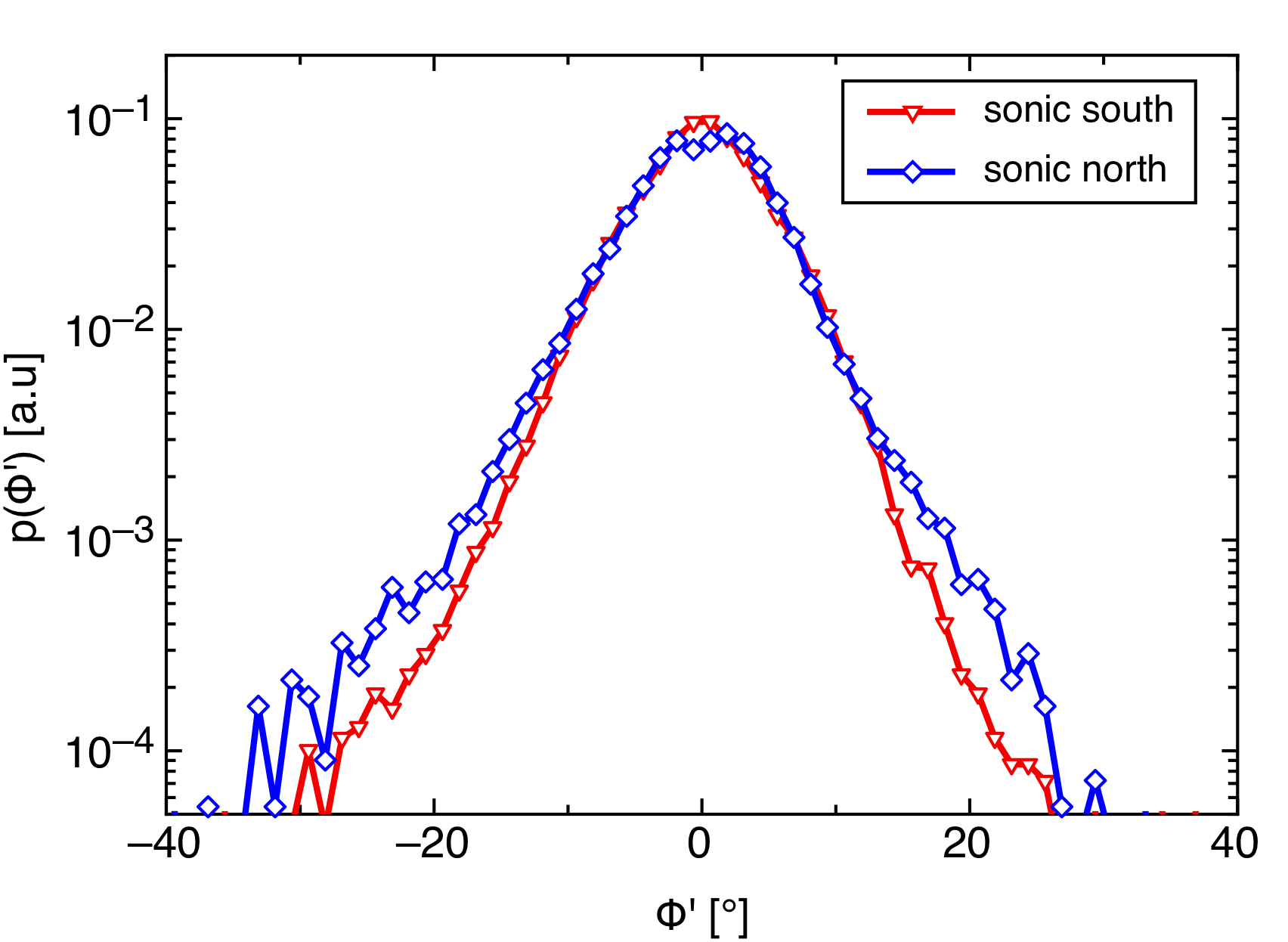}%
  \begin{picture}(0,0)%
    \put(-90,113){(a)}
    \put(-90, 50){(b)}
  \end{picture}%
  \caption{(a) PDF of the wind direction fluctuations measured with
    the 3D ultrasonic anemometer in northward (blue) and southward
    (red) orientation at about 10\,m/s inflow velocity. (b) PDF of the
    wind direction fluctuations measured with the 3D ultrasonic
    anemometer in northward (blue) and southward (red) orientation at
    about 20\,m/s inflow velocity.}
  \label{fig:PDFAnemometers}
\end{figure}
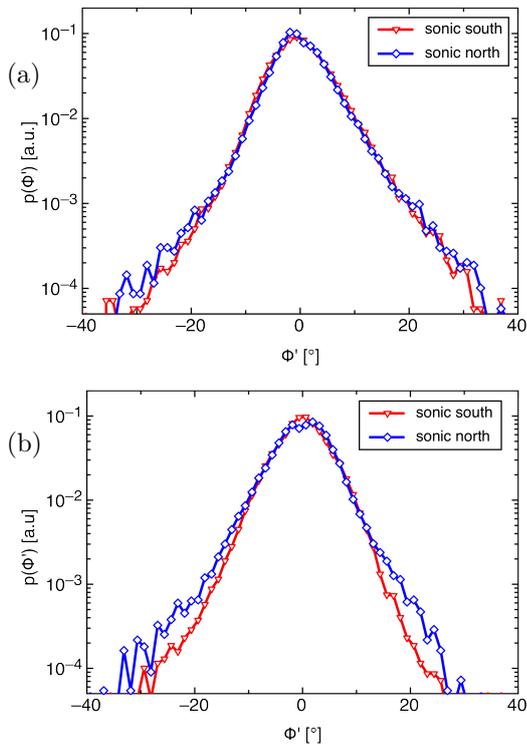

Besides the comparison of the wind speed measurements, it is also
worth comparing the wind directions measured with the 3D ultrasonic
anemometer in different orientations. For this purpose the PDFs of the
wind direction fluctuations $\phi'(t) = \phi(t) - \langle
\phi(t)\rangle$ are shown in Fig.~\ref{fig:PDFAnemometers} for an
inflow velocity of about 10\,m/s (a) and about 20\,m/s (b). The
comparison of the two different orientations for both wind speeds
gives evidence of the increased probability for the occurrence of
larger direction fluctuations for the northward alignment (0$^\circ$)
of the anemometer. This result is expected, since a supporting rod is
located upstream the measuring volume and thus vortex shedding causes
wake effects. The self-induced effect is more pronounced for the
higher velocity in Fig.~\ref{fig:PDFAnemometers}(b). Additionally,
the measured mean wind speed of 16.2\,m/s for northward alignment is
reduced by approximately 1.4\,m/s compared to the southward
orientation. Considering the geometry of the ultrasonic anemometer an
increasing systematic error in the measurements can be concluded for
inflow from 0$^\circ$, 120$^\circ$ and 240$^\circ$, respectively.

To conclude this subsection, the active grid offers the generation of
turbulent flows with similarly intermittent statistics to atmospheric
wind as was shown in \cite{Knebel2011}. Especially promising in this
context is also the generation of reproducible flow structures in a
deterministic sense, as it has been shown in this subsection. Besides
anemometer testing and characterization, also the interaction of
typical structures like airfoils or model wind turbines with
reproducible flow fields shall be investigated in the future.
It should be noted that the active grid offers also further
applications for wind energy research, e.g., generating shear flows
\cite{Cekli2010} or special structures like low level jets
and large scale intermittency \cite{Good2011}.


\subsection{Impact on loads and power generation of wind energy
  converters}
\label{sec:loads}

The turbulent nature of the wind with its characteristic intermittency
can be expected to have a major impact on wind energy converters.
These machines  operate in the turbulent atmospheric boundary layer
and are thus permanently exposed to turbulent wind fields.
Among the many aspects of a complete WEC system, we choose here the
torque and thrust at the rotor as central force and load quantities to
investigate the influence of the turbulent inflow characteristics on
WECs.

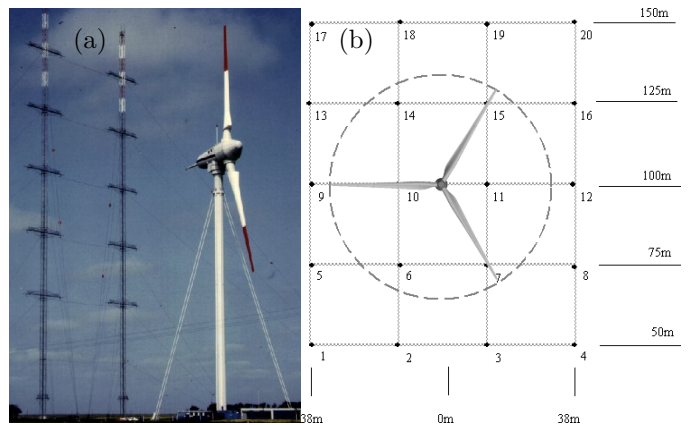
\begin{figure}
  \centering
  \hspace*{\fill}%
  \includegraphics[height=55mm]{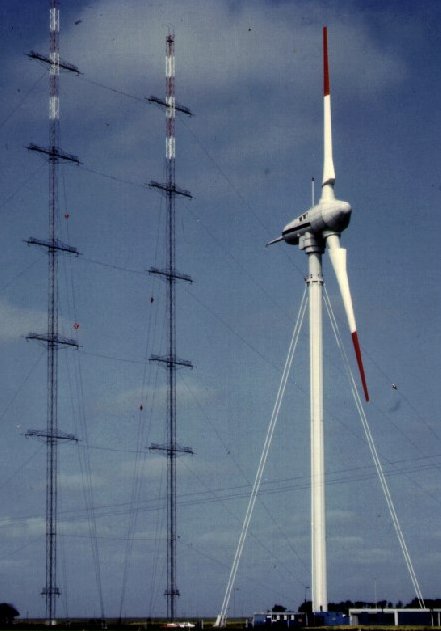}%
  \hspace*{\fill}%
  \includegraphics[height=55mm]{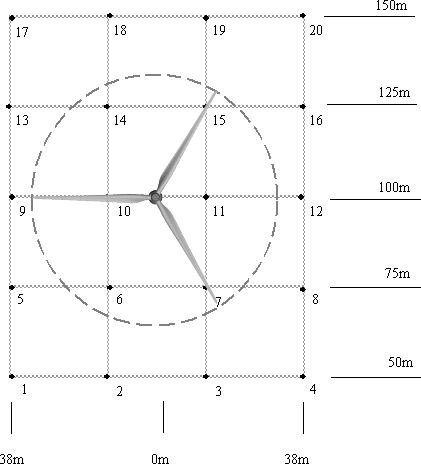}%
  \hspace*{\fill}%
  \begin{picture}(0,0)%
    \put(-103,64){(a)}%
    \put( -58,64){(b)}%
  \end{picture}%
  \\
  \caption{(a) GROWIAN test site of the 1980's \cite{Guenther1998} and
    (b) layout of the wind measurement setup, compared to the position of
    the rotor. }
  \label{fig:growian}
\end{figure}
\begin{figure}
  \centering%
  \hspace*{\fill}%
  \includegraphics[width=0.45\linewidth]{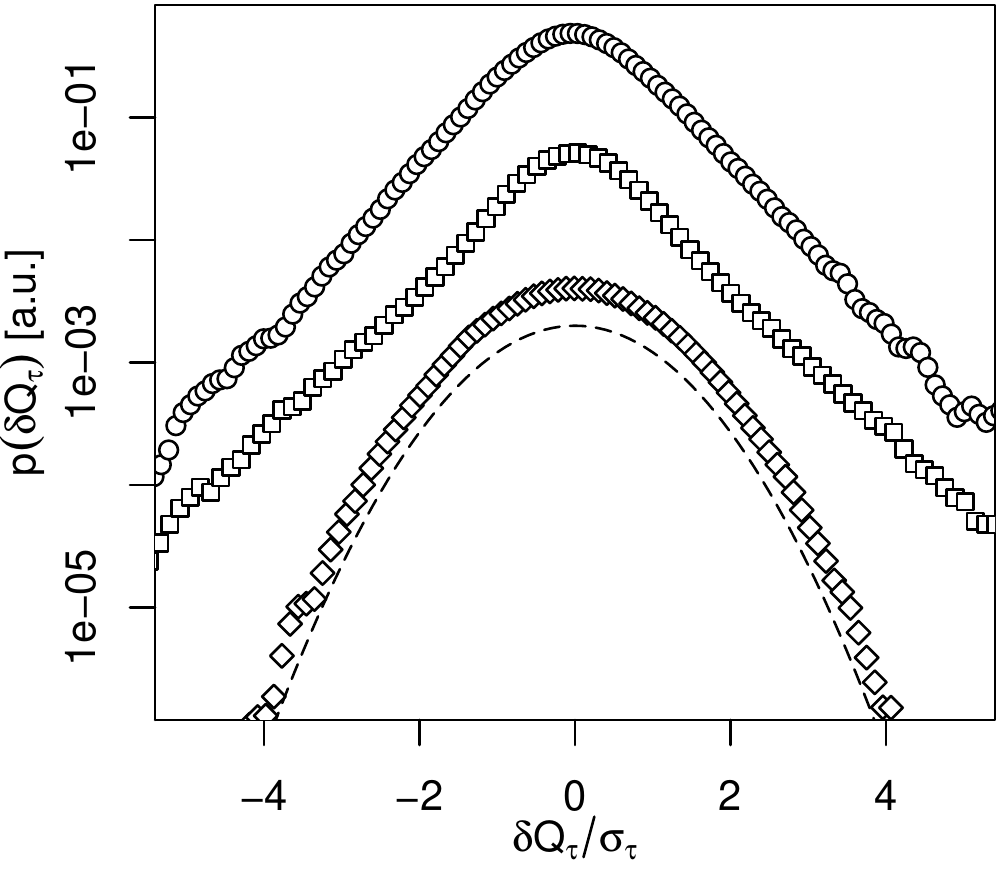}%
  \begin{picture}(0,0)%
    \put(-40,40){(a)}%
  \end{picture}%
  \hspace*{\fill}%
  \includegraphics[width=0.45\linewidth]{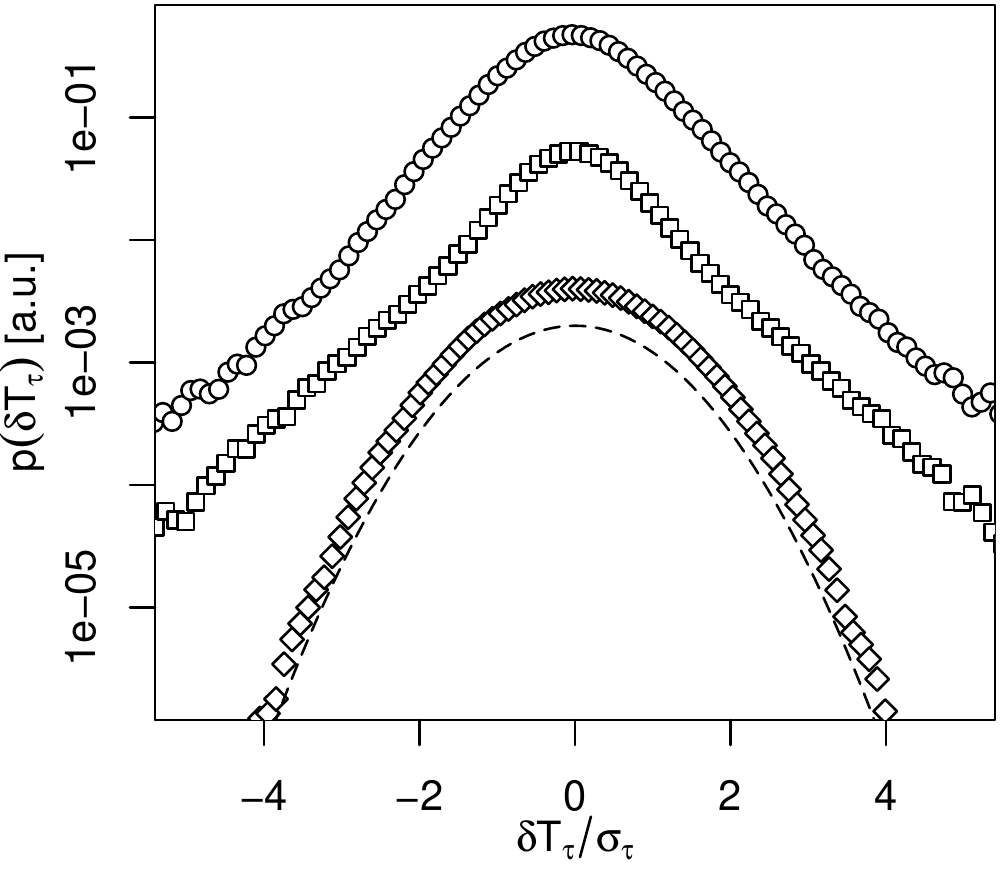}%
  \begin{picture}(0,0)%
    \put(-40,40){(b)}%
  \end{picture}%
  \hspace*{\fill}%
  \caption{PDFs of %
    (a) rotor torque increments, $\delta Q_\tau=Q(t+\tau)-Q(t)$, and %
    (b) rotor thrust increments, $\delta T_\tau=T(t+\tau)-T(t)$, %
    derived from simulations of the WIndPACT~1.5\,MW virtual turbine.
    Wind inflow data were used from the GROWIAN wind field measurement
    (circles, top), as generated by a stochastic intermittent wind
    field model (squares, middle), and from the output of the TurbSim
    software using the Kaimal model (diamonds, bottom). Gaussian PDFs
    are shown as dashed lines for comparison. Time scale $\tau$ is 1\,s
    for all presented PDFs. 
    Results of (a) have been published in \cite{Muecke2010}.
  }
  \label{fig:Muecke}
\end{figure}
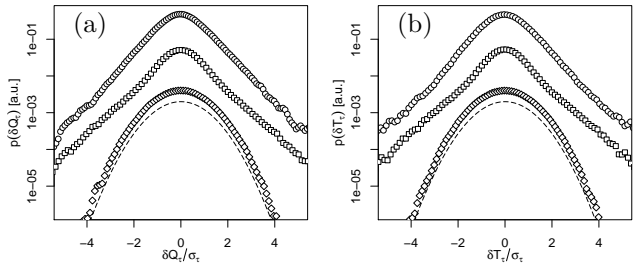

Numerical studies have been performed using the FAST code by NREL
\cite{Jonkman2005} together with the WindPACT 1.5\,MW virtual
turbine \cite{Malcolm2002}. The unique wind field measurement of the
German GROWIAN campaign \cite{Koerber1988, Guenther1998} was used as
inflow condition. Additionally the Kaimal model of the TurbSim
software \cite{Kelley2011} and a newly developed model based on
stochastic processes
\cite{Kleinhans2006, Kleinhans2008} were used for comparison.
In a first analysis, the statistics of rotor torque increments were
derived and compared \cite{Muecke2010}.
It was found that the typical intermittency of the wind transfers to a
similar intermittency in the rotor torque when the measured wind field
is used as input, see \figref{fig:Muecke}(a). The Gaussian increment
statistics of the Kaimal wind field, on the other hand, lead to
Gaussian torque statistics, supporting the statements above on the
implicit Gaussian assumptions in the IEC standard \cite{IEC2005a}. The
intermittent synthetic wind field results in much more realistic
torque statistics.

Here we present additional results of the rotor thrust forces from the
identical numerical setup as above. In \figref{fig:Muecke}(b) thrust
increments obtained under the same inflow conditions as in part (a)
are compared at $\tau=1$\,s. 
We find that the measured wind field leads to a highly intermittent
torque PDF, which is well reproduced by the stochastic wind field
model. The Kaimal wind field leads to purely Gaussian thrust
increment PDFs and thus underestimates extreme thrust loads.

Both torque and thrust form a complete decomposition of forces at the
nacelle of a WEC. With these results we can therefore conclude that
intermittency is a general property of the wind-generated forces at
the WEC. Moreover, thrust forces appear at all types of buildings and
structures which are exposed to the wind. This emphasizes again the
relevance of our results.

As we observe intermittent statistics of the rotor forces at WECs, it
can be expected that the generated electrical power signal is also
influenced by this property. In \litref{Gottschall2007} it could be
demonstrated that actually the electrical power signal possesses
similarly intermittent dynamics as the wind. These results are
confirmed and extended in this paper, see \secref{sec:modeling}.

From our results we see that the use of Gaussian wind fields in WEC
simulation and design must lead to a dramatic under-estimation of
extreme torque, thrust, and thus load changes, which can be comparable
to \figref{fig:intermittency}. Most interestingly these load features
are not grasped by the characterization based on standard rain flow
counting methods.
The intermittent statistics of wind and loads are then transferred
into an evenly intermittent signal of the electrical power output.
With a growing share of wind energy fed into the electrcal grid, the
high probabilities of extreme changes will have a more and more
significant impact on the grid.
Another important aspect also becomes clear from these results. Short
time fluctuations are not smoothed out by, e.g., some possible spatial
averaging or rotational inertia of the rotor. In contrast the nature
of the turbulent wind amplifies significantly the effects of the short
time fluctuations. For further developed WECs of the future, e.g., the
10\,MW class, these turbulent properties have to be considered even
more seriously.

To conclude, with the statistics of rotor torque and thrust increments
we achieve a complete characterization of the forces at the rotor of
WECs. We find that the intermittency of the wind is transferred to the
generator by the torque, as well as to the complete structure of the
WEC by the thrust forces. The resulting electrical power output of
these intermittent input forces will be subject of the following
section.

\section{Characterisation and modeling of wind-generated 
  electrical power}
\label{sec:modeling}

\begin{figure}[ht]
  \centering
  \includegraphics[height=50mm]{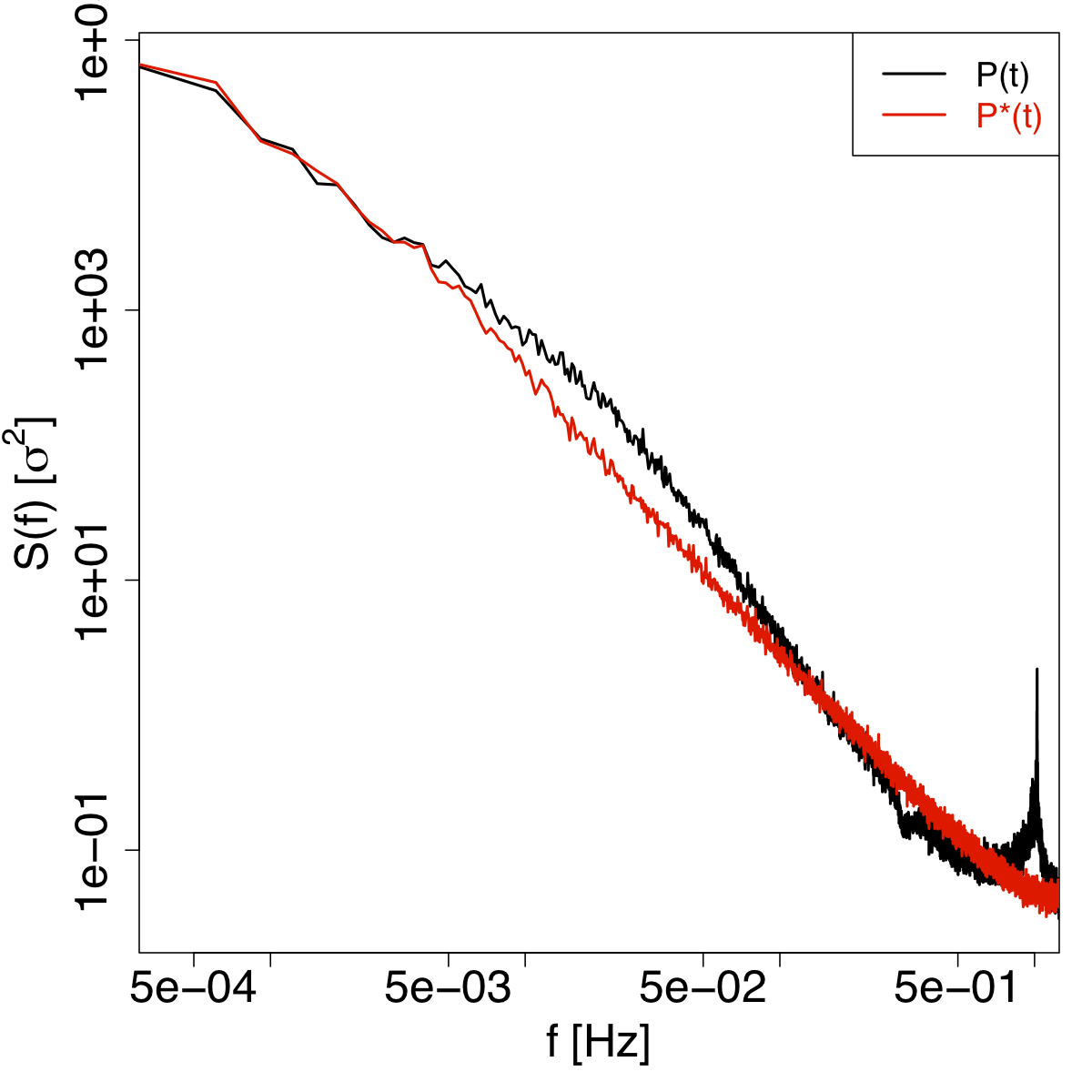}\\
  \includegraphics[height=50mm]{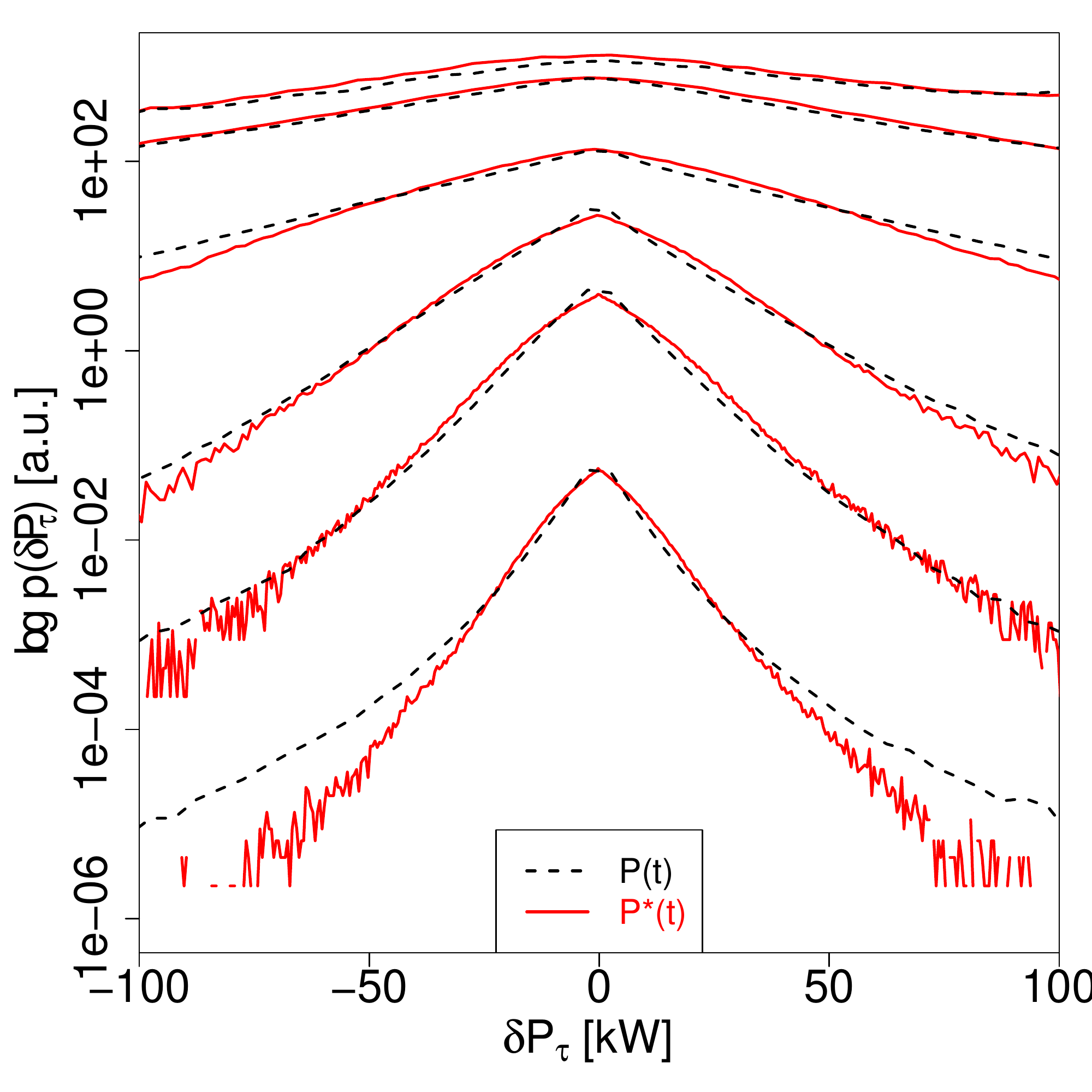}%
  \begin{picture}(0,0)%
    \put(-70,110){(a)}%
    \put(-70, 50){(b)}%
  \end{picture}%
  \caption{(a) Power spectral density $S(f)$ of $P$ (black) and
    $P^{\star}$ (red). $S(f)$ is normalized by the variance $\sigma^2$
    of the time series. (b) PDF of power increments $\delta P_{\tau}$
    for various scales $\tau=(0.4, 0.8, 1.6, 6.4, 25.6, 26214.4)$\,s
    from bottom to top, curves are shifted vertically for clarity. The
    solid red lines are estimated from $P^{\star}$ and the black
    dashed lines from $P$. The vertical axis is represented using a
    logarithmic scale. 
  }
  \label{fig:power_stat}
\end{figure}
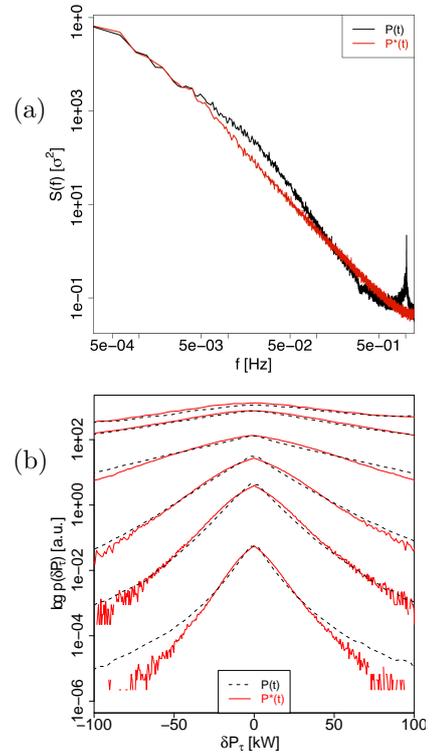

Throughout the previous sections, it could be demonstrated that the
intermittent nature of atmospheric turbulence influences all stages of
the wind energy conversion process. The wind imposes its intermittent
character upon all relevant quantities, such as wind speeds,
mechanical forces and torques, and electrical power output of WEC. 
To achieve progress in wind power analysis and modeling, alternative
methods to those defined in the current standards \cite{IEC2005a} are
needed. Here we present a method adapted to the intermittent character
of atmospheric wind. 
The highly dynamical character of the power conversion process and the
intermittent properties of the electrical power output have been shown
before \cite{Anahua2007, Gottschall2008, Gottschall2007}. They are
confirmed by the results of this section,
cf.~\figref{fig:power_stat}(b).

Future sustainable energy supply systems will have to manage large
shares of fluctuating energy sources such as wind energy. This
constitutes a demand of properly modeling the high frequency dynamics
of these energy conversion systems, including their nonlinear and
intermittent properties. As an alternative to highly detailed, complex
(and thus computationally expensive) models the usage of stochastic
models has recently found a wide range of applications
\cite{Friedrich2009, Friedrich2011}. Generally spoken, for complex
systems this approach tries to separate only few crucial
variables\footnote{so-called order parameters, following
  \cite{Haken1983}} from a larger number of less important ones, which
are then replaced by simple noise terms. By doing so, in many cases
simply structured models are obtained which nevertheless cover the
essential dynamics of the system under investigation. Their parameters
can be derived directly from measurement data,
cf.~\cite{Friedrich2011}.

Recently the Langevin Power Curve (LPC) has been developed as a
dynamical power characteristic of WECs \cite{Anahua2007,
  Gottschall2008, Waechter2010}. By this method, for each wind speed
the stable (or attractive) fixed points of the energy conversion
dynamics are obtained. Together they constitute the LPC of the WEC,
which provides important information, complementary to the current
industry standard IEC 61400-12-1 \cite{IEC2005a}. 

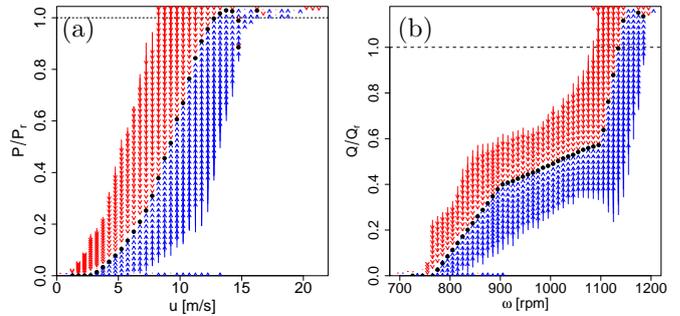
\begin{figure}
  \centering
  \hspace*{\fill}%
  \includegraphics[width=0.49\linewidth]{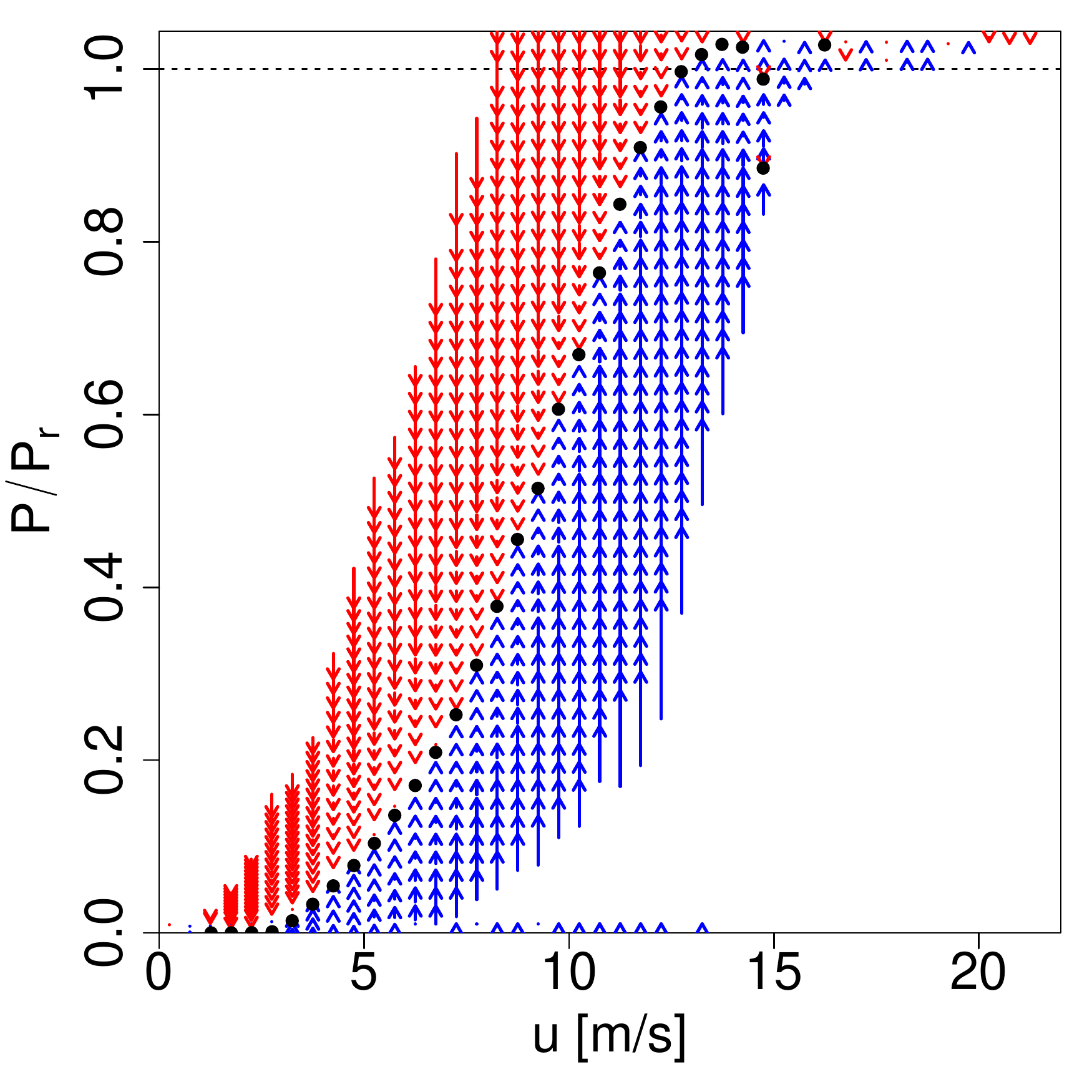}%
  \hspace*{\fill}%
  \includegraphics[width=0.49\linewidth]{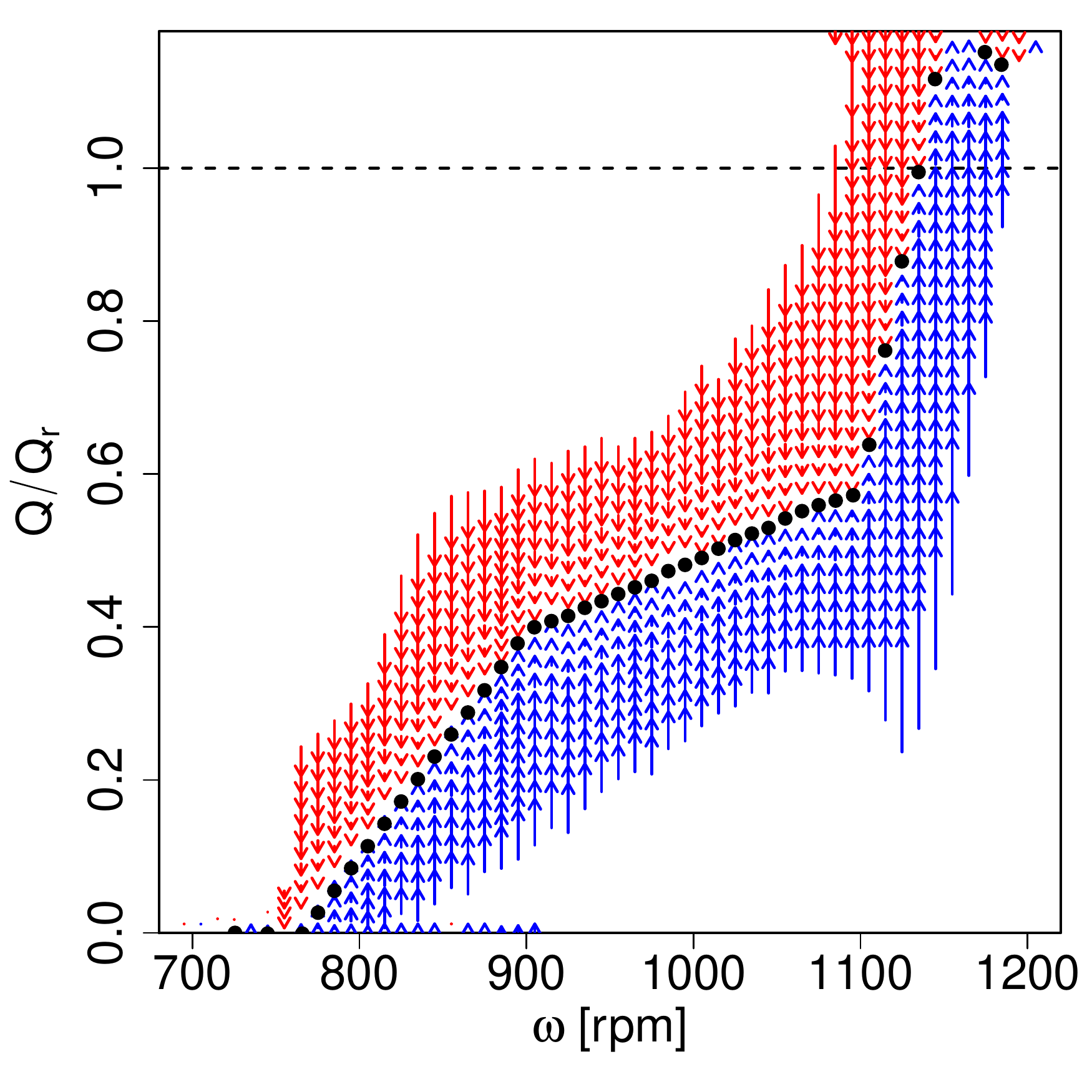}%
  \hspace*{\fill}%
  \begin{picture}(0,0)%
    \put(-104, 49){(a)}%
    \put( -47, 49){(b)}%
  \end{picture}%
  \caption{%
    Drift fields $D^{(1)}$ derived from WEC measurements,
    represented qualitatively using arrows. The local value of the drift
    $D^{(1)}$ is represented by the direction and length of each
    arrow, while the color also indicates the direction of the drift.
    The black dots represent the stable fixed points where
    $D^{(1)}=0$.
    (a)~Drift field $D^{(1)}(P;u)$ of electrical power output vs.\
    wind speed. Here the black dots repesent the Langevin Power Curve
    \cite{Milan2010b}. 
    (b) Drift field $D^{(1)}(Q;\omega)$ of rotor torque vs.\ angular
    velocity. For this 
    analysis the torque was obtained from available measurements of
    angular velocity and electrical power output.
    Power and torque are normalized by their rated values $P_r$ and
    $Q_r$, respectively. 
  }
  \label{fig:drift}
\end{figure}
Here we extend this concept and propose a stochastic model of the
electrical power output of WEC, driven by the inflowing wind.
The structure of the model aims towards the high-frequency dynamics of
the conversion from wind speed to electrical power. The power output
$P(t)$ is modeled by a stochastic differential equation, namely the
Langevin equation \cite{Risken1984}
\begin{equation}
  \label{eq:langevin}
  \frac{d}{dt}P(t)=D^{(1)}(P;u)+\sqrt{D^{(2)}(P;u)}\cdot \Gamma(t) \, .
\end{equation}
Here $D^{(1)}(P;u)$ and $D^{(2)}(P;u)$ are the drift and diffusion
matrices, respectively, and $\Gamma(t)$ is a Gaussian white noise with
mean value $\langle \Gamma(t) \rangle=0$ and variance $\langle
\Gamma^2(t) \rangle=2$. The drift and diffusion matrices
$D^{(1)}(P;u)$ and $D^{(2)}(P;u)$ can be estimated directly from
measurement data (sampled at the order of $1\mathit{Hz}$), as follows
\cite{Risken1984, Anahua2007}
%
\begin{multline}
  \label{eq:kmc}
  D^{(n)}(P;u)=\\
  \frac{1}{n!} \lim_{\tau \to 0} \frac{1}{\tau}
  \left\langle \left[P(t+\tau)-P(t)\right]^n|P(t)=P;u(t)=u\right\rangle \, ,
\end{multline}
%
where $\langle\cdot|\cdot\rangle$ denotes conditional ensemble
averaging. Note that $u(t)$ is thus conditioned on a fixed value $u$.
An illustration of $D^{(1)}$ is given in \figref{fig:drift}(a).
The drift field (or matrix) $D^{(1)}$ represents the dynamical map of the
conversion process $u(t) \rightarrow P(t)$. 
The power production is attracted towards the stable fixed points
$P(u)$, while the turbulent wind fluctuations continuously drive the
system away. This corresponds to the structure of the Langevin
equation (\ref{eq:langevin}), where the drift field $D^{(1)}$
represents the attractor displayed in \figref{fig:drift}(a), and where
the diffusion field $D^{(2)}$ quantifies additional, random
fluctuations stemming from turbulence. A second example of a drift
field is presented in \figref{fig:drift}(b). Here the dynamics of the
torque depending on the rotational speed is displayed, showing that
the method of the LPC is not limited to the dynamics of power output.
It rather defines a general mathematical tool for analyzing the
dynamics of complex, noisy, and noise-driven systems.

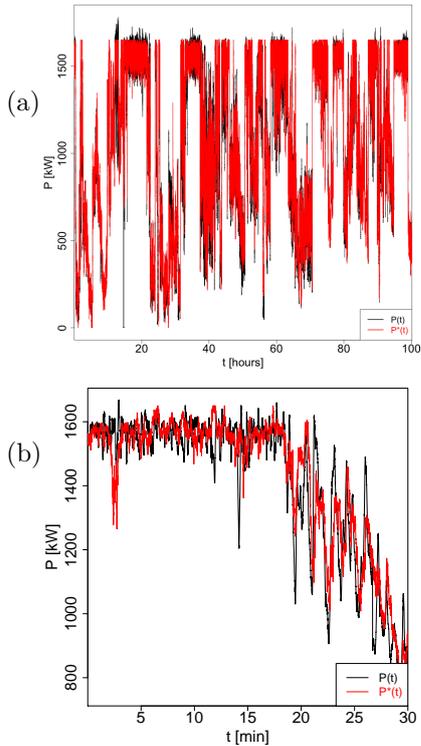
\begin{figure}
  \centering
  \includegraphics[height=50mm]{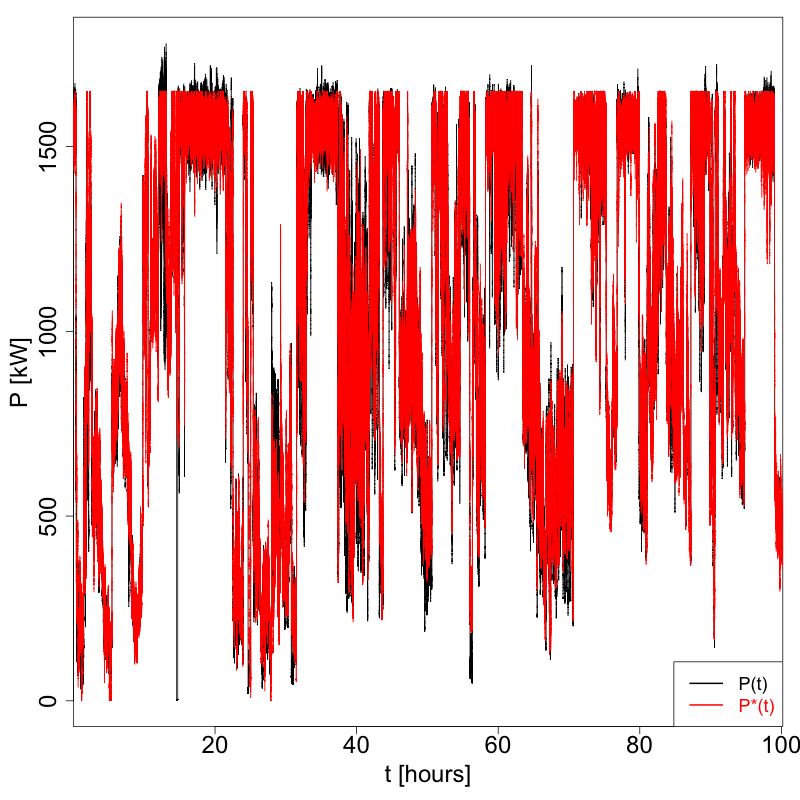}\\
  \includegraphics[height=50mm]{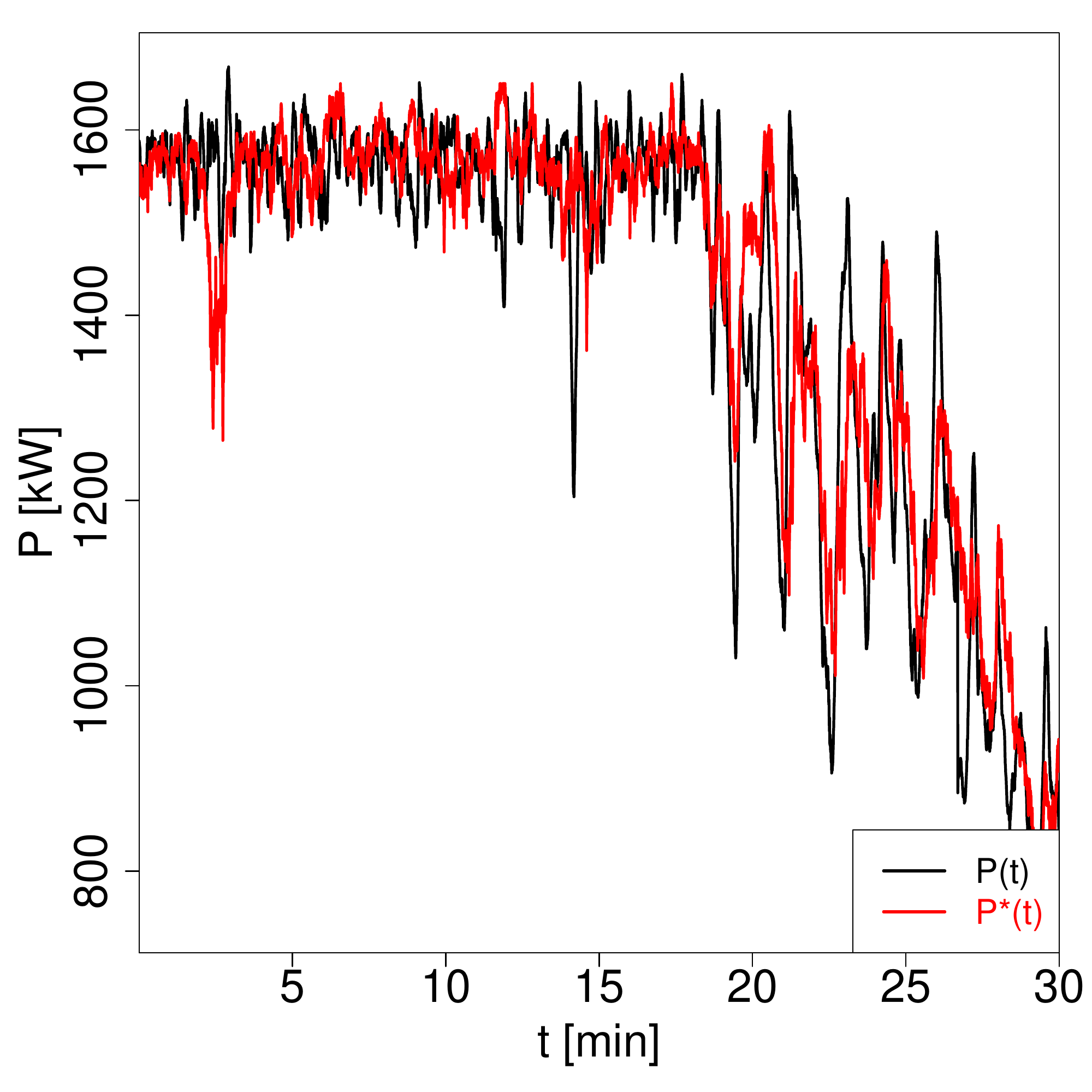}%
  \begin{picture}(0,0)%
    \put(-70,110){(a)}%
    \put(-70, 50){(b)}%
  \end{picture}%
  \caption{Comparison of $P(t)$ (black) and $P^{\star}(t)$ (red) for
    (a) 100\,h and (b) 30\,min at sampling frequency 2.5\,Hz.}
  \label{fig:series}
\end{figure}
While only the drift field with its fixed points as shown in
\figref{fig:drift} defines the LPC, the Langevin equation
\eqref{eq:langevin} uses both the drift and diffusion matrices.
Once these
have been estimated from a dataset, they can be used in the Langevin
equation to simulate the power output $P(t)$ for any given wind speed
$u(t)$.

As a first hint for the validity of the stochastic model\footnote{%
  The results of the stochastic model are presented here for a dataset
  obtained from a numerical model, see details in \litref{Milan2011a}.
  Similar results were obtained for various measurement datasets, for
  which the results were surprisingly good, even though the noise by
  the turbulent inflow will not fulfill the conditions of a Langevin
  noise.}, a visual comparison
of $P$ and $P^{\star}$ is given in \figref{fig:series}.
A statistical comparison is performed based on power spectral
densities in \figref{fig:power_stat}(a). The overall shape of the
spectrum is reproduced by the stochastic model. The most striking
deviation is observed at frequency $f \approx 1$\,Hz, which
corresponds to a $1$\,Hz oscillation in the measured time
series $P(t)$. We believe this is due to the interaction of the tower
with the rotating blades of the WEC\footnote{It should be noted that not
  only the tower interaction can generate periodic fluctuations in the
  order of 1\,Hz. Additional contributions from the rotating
  rotor are not identified here.}, 
which happens in the same frequency range. 
As our model is based on a first-order stochastic differential
equation \eqref{eq:langevin} it can in principle not reproduce a
single-frequency oscillation like this. As a future improvement, a
suitable oscillation could be added to the model to compensate for
this limitation.\footnote{Here it should be noted also that in a typical
measurement setup with 1\,Hz sampling frequency this oscillation would
not be visible, making the validation of such a model extension
impossible in most cases.}
Figure \ref{fig:power_stat}(b) illustrates the ability of the model to
reproduce the intermittency of the original power signal. For example,
changes in power production up to $\pm 100$\,kW within $0.8$\,s occur.
This aspect is especially important on shorter time scales when fast
wind gusts are converted into fast power gusts. Moderate deviations
are observed on the shorter time scales, where the model behaves less
intermittent than desired. Overall, the stochastic model obtains 
valuable results as spectral properties and intermittency are mostly
well reproduced. Its fast and flexible structure makes the stochastic
method useful for a realistic modeling of single WECs, wind farms and
wind farm clusters (as will be shown elsewhere).

It should nevertheless be noted that the purpose of this model is not
to compete with current full-featured multibody models of WECs, such
as FAST \cite{Jonkman2005} or FLEX \cite{Oye1999}. In contrast to
these, the proposed model is highly simplified and does not aim at a
detailed model of the processes inside a WEC. It rather constitutes an
effective model of a WEC as a whole, considering only the wind as
input and a single output quantity, such as electrical power or
torque. While the descriptive power of our model in this sense is
limited, it benefits from high numerical efficieny as well as easy and
straightforward parameter retrieval.

We can conclude that the stochastic model is able to capture
consistently the dynamics of power or torque, depending on wind speed
or rotational frequency, respectively. Once the parameters have been
estimated from measurements, for simulations only wind speed time
series are required as input. In the results especially the typical
intermittency is reproduced.


\section{Outlook: n-point statistics and forecasting}
\label{sec:outlook}

Up to this point our analysis, modeling, and experiments concerning
atmospheric turbulence were concentrating on increments and
fluctuations of relevant signals, e.g., wind speeds, torque, and
electrical power. These statistics can be obtained in a
straightforward way and it has been demonstrated that they are of high
relevance.

Based on recent results presented in \secref{sec:analysis} we
are presently working on the decomposition of atmospheric turbulence into
components which behave similar to homogeneous isotropic turbulence. 
For this case, in the recent years a framework of stochastic analysis
and characterization has been developed, which grasps the complete
statistical properties of turbulence in terms of arbitrary $n$-scale
joint probabilities \cite{Friedrich1997b, Renner2001, Friedrich2009,
  Friedrich2011}. Moreover, a method has been presented for the
multiscale reconstruction of real turbulent time series
\cite{Nawroth2006, Stresing2010a}.

From the combination of these work topics in the future we hope to
make adcvances towards a multipoint description framework for
atmospheric turbulence. This would mean also a proper characterization
of multiscale joint probabilities, which correspond to coherent
structures in wind fields. As a consequence, also multiscale
short-time forecasting of wind events, such as gusts and clustering
effects, could be possible with only little computational effort,
compare \litref{Nawroth2006}.


\section{Conclusions}

In this contribution, fundamental as well as applied aspects have been
presented for advanced statistics, modeling and experiments concerning
the unique small-scale intermittency features of atmospheric
turbulence and their impact on the wind energy conversion process.
Based on the fact that atmospheric turbulence
exhibits these complex statistical properties especially in wind speed
increment and fluctuation PDFs, we proposed an advanced
characterization. Similar to the modeling and characterization of wind
data in \cite{Boettcher2007}, here we achieve a proper description and
modeling of atmospheric fluctuation PDFs.
First successful attempts have been made towards the decomposition of
atmospheric turbulence into subsets which share the properties of
homogeneous isotropic turbulence.
Based on these results a more realistic, yet still practicable,
characterization of atmospheric wind becomes possible, overcoming the
implicit assumption of Gaussianity in current methods and standards.

As turbulence, in principle, has to be considered an open problem of
classical physics, experimental work is indispensable. For the
reproduction of crucial properties of atmospheric flows the active
grid offers extreme flexibility in the design and control of wind
tunnel experiments.
Investigations of the reproducibility of flow structures have been
presented as well as the application to testing and characterization
of anemometer properties.

The impact of atmospheric turbulence on wind energy conversion is
investigated in two separate aspects. Numerical modeling of turbines
shows evidence that the intermittent nature of the wind leads to high
probabilities of extreme load changes in both torque and thrust. These
results completely characterize the forces induced by the rotor to the
drive train of a WEC.
Accordingly intermittent statistics are found in the electrical power
output of both numerical models and real turbines, which constitutes
the input to the electrical grid.
An extended approach can model the power output of a WEC based on wind
measurement data, reproducing properly the statistical properties of
the real power output signal.

The common assumption of Gaussian wind statistics necessarily neglects
these intermittency effects. We propose that more realistic
statistical description and modeling of atmospheric turbulence is
necessary to advance not only in science but also in engineering as
well as in manufacturing and operation of WECs.

Stochastic methods are shown as appropriate tools to characterize the
dynamics and intermittency of the wind energy conversion process in
more details, and thus provide valuable, additional information to
standard methods. 
The results presented show that these methods have a large potential
of application for further aspects of wind energy.


\bibliographystyle{tJOT_mw}
\bibliography{Wind+Turbulence}

\end{document}